\documentclass[
]{revtex4-2}
\usepackage{graphicx} 
\usepackage{comment}
\usepackage{dsfont}
\usepackage{color}
\usepackage{bm} 
\usepackage{hyperref}

\usepackage[utf8]{inputenc}
\usepackage{amsmath}
\usepackage{amssymb}
\usepackage[a4paper, total={6.6in, 8.8in}]{geometry}
\usepackage{graphicx}
\usepackage{hyperref}
\usepackage{float}
\usepackage{braket}
\usepackage{adjustbox}
\usepackage{color}
\usepackage{framed} 
\usepackage{url}
\usepackage[normalem]{ulem}

\begin{document}

\begin{abstract}

Common wisdom dictates that physical systems become less 
ordered when heated to higher temperature. However, several 
systems display the opposite phenomenon and move to a more 
ordered state upon heating, e.g. at low temperature piezoelectric 
quartz is paraelectric and it only becomes piezoelectric 
when heated to sufficiently high temperature. 
The presence, or better, the re-entrance of unordered phases 
at low temperature is more prevalent than one might think. 
Although specific models have been developed to understand 
the phenomenon in specific systems, a universal 
explanation is lacking. Here we propose a universal simple 
microscopic theory which predicts the existence of two 
critical temperatures in inhomogeneous systems, where the 
lower one marks the re-entrance into the less ordered phase. 
We show that the re-entrant phase transition is caused 
by disorder-induced spatial localization of the zero-mode on 
a finite, i.e. sub-extensive, region of the system. 
Specifically, this trapping of the zero-mode 
disconnects the fluctuations of the order parameter 
in distant regions of the system, thus triggering the
loss of long-range order and the re-entrance into the 
disordered phase. 
This makes the phenomenon quite universal and robust to the 
underlying details of the model, and explains its ubiquitous 
observation. 
\end{abstract}

\title{Re-entrant phase transitions induced by localization of zero-modes}

\author{Flaviano Morone}
\affiliation{Department of Physics, New York University, New York, NY, USA}
\author{Dries Sels}
\affiliation{Department of Physics, New York University, New York, NY, USA}
\affiliation{Center for Computational Quantum Physics, Flatiron Institute, New York, NY, USA}

\maketitle

\begin{figure}[h!]
\includegraphics[width=0.75\textwidth]{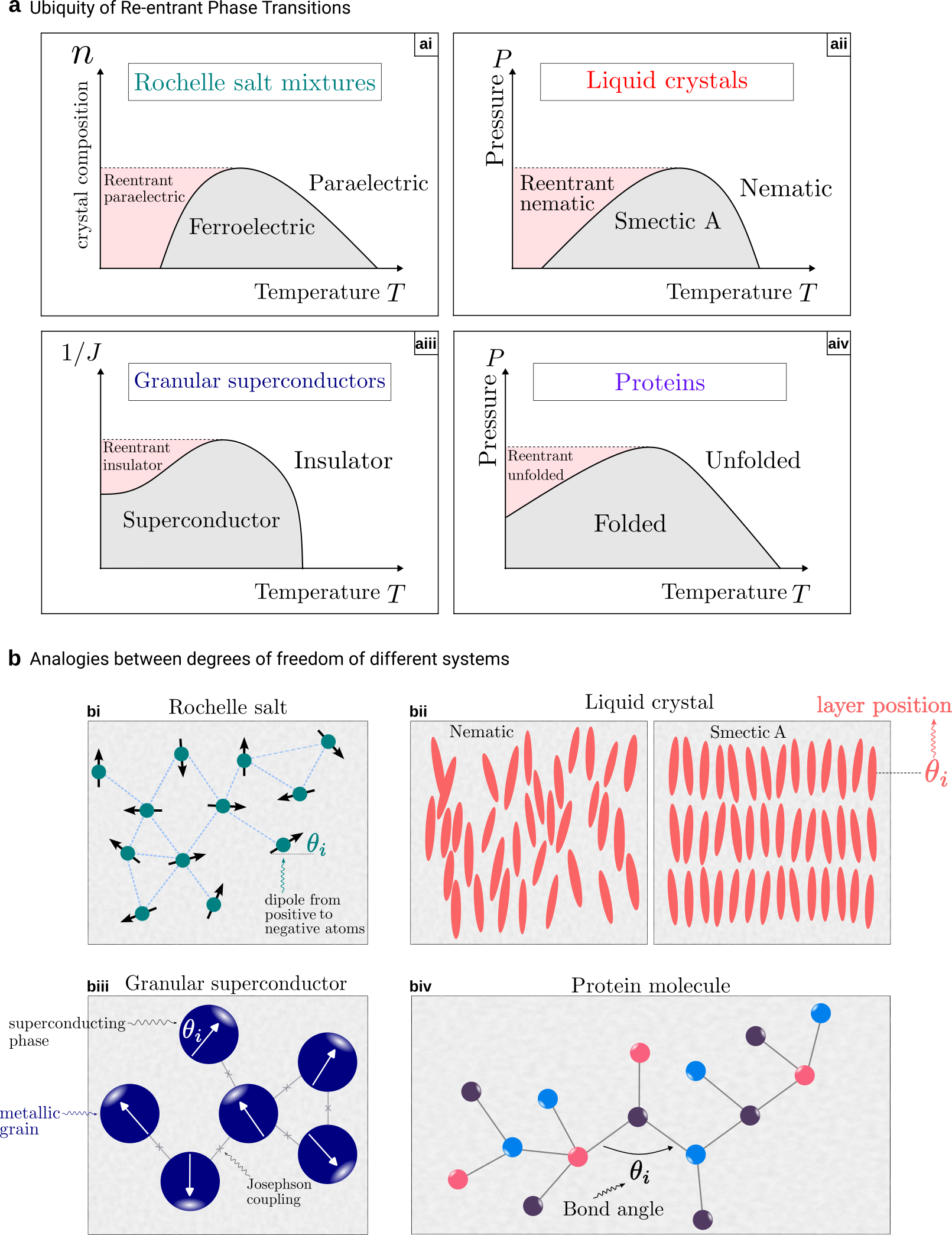} 
\caption{{\bf Ubiquity of reentrance and universal microscopic 
model.}
\textbf{a,} ({\it ai}) Phase-diagram of Rochelle 
salt showing a re-entrant paraelectric phase at low 
temperature~\cite{valasek, cladis2}; ({\it aii})
re-entrant nematic phase in liquid crystals~\cite{cladis1, cladis2}; 
({\it aiii}) re-entrant insulating phase in granular superconductors~\cite{simanek2,efetov}; 
({\it aiv}) re-entrant unfolded (or denatured) state 
in protein folding~\cite{kauzmann}. 
\textbf{b,} ({\it bi}) The variable $\theta_i$ denotes 
the direction of the dipole from positive atoms to 
negative atoms in Rochelle salt; ({\it bii}) the position 
of the layers in a liquid crystal; ({\it biii}) the 
superconducting phase of metallic grains 
in disordered Josephson arrays; ({\it biv}) 
the bond angle of amino acids within a protein. 
}
\label{fig:fig1}
\end{figure}

Rochelle salt began to excite interest since the Curie 
brothers discovered its fascinating piezoelectric 
properties in 1880. Even more remarkable was the 
discovery~\cite{valasek}, forty years later, that 
Rochelle salt had two Curie temperatures: 
$T_c^{(1)}= 24^{\rm o}C > T_c^{(2)} = -18^{\rm o}C$. 
Above $T_c^{(1)}$ and below $T_c^{(2)}$ Rochelle salt is 
paraelectric (there is no spontaneous polarization) and 
ferroelectric in between them (see Fig.~\ref{fig:fig1}ai). 
It is, perhaps, the first known case of a re-entrant 
phase transition ever observed in nature~\cite{cladis2}. 
Subsequently, re-entrant transitions have been discovered 
in several physical and biological systems, including the 
insulator$\to$superconductor$\to$insulator transition
in granular superconductors~\cite{simanek2, efetov, blatter},  
the nematic$\to$smectic A$\to$nematic transition in 
liquid crystals~\cite{cladis1, cladis2}, and the 
unfolded$\to$folded$\to$unfolded transition in protein folding~\cite{kauzmann}, to name a few examples (see 
Figs.~\ref{fig:fig1}aii,aiii,aiv). 

The phenomenon of re-entrance has, of course, generated 
several theoretical ideas, each in its own way successful 
on some scale in describing 
observations~\cite{vaks, berker, prost, simanek1, simanek2, efetov}. 
On the other hand, in models where it is found, it occurs 
for a small range of the parameters and then completely 
disappears in different dimensions~\cite{vaks, efetov}. 
More importantly, the general physical mechanism of 
re-entrance and its robustness remains unexplained. Here 
we suggest a simple universal theory of re-entrant phase 
transitions, which sheds light on the physical mechanism 
causing the re-entrance of the less ordered phase at low 
temperature. Specifically, 
we show that the spatial localization of the Goldstone zero-mode 
leads to the loss of long range order as the temperature is lowered. 
A simple variational approximation to the problem elucidates 
the non-perturbative nature of this effect. 
To appreciate the ubiquity of re-entrant phases, we sketch 
in Fig.~\ref{fig:fig1} the phase diagrams for a variety 
of real systems, including ferroelectric mixtures (Fig.~\ref{fig:fig1}ai), 
liquid crystals (Fig.~\ref{fig:fig1}aii), granular 
superconductors (Fig.~\ref{fig:fig1}aiii), and protein 
molecules (Fig.~\ref{fig:fig1}aiv), all displaying a prominent 
re-entrant phase at low temperature. 
The effective degrees of freedom of these systems are, in fact, 
in close analogy to each other, in that the nematic-smectic A 
transition in a liquid crystal is isomorphous to the phase-locking 
transition in an assembly of superconducting grains and to the 
ferrocoherent transition in Rochelle salt~\cite{valasek, scott, 
pellan, deutscher, degennes}. 
Impurities are an essential, and often unavoidable,  
ingredient making up these systems, that are modeled 
by random vectors coupled to the order parameter  
representing local random magnetic fields in superconductors 
and ferromagnetic materials~\cite{maple}, local twists 
and bend deformations in liquid crystals~\cite{degennes} 
and protein molecules~\cite{dill}. The statistical mechanical 
model that captures all these systems at once is described 
by the Hamiltonian:
\begin{equation}
{\mathcal H} = -\frac{1}{2}\sum_{i,j=1}^N J_{ij}A_{ij}\cos(\theta_i-\theta_j) - 
\sum_{i=1}^{N}H_i\cos(\theta_i-\phi_i)\ ,
\label{eq:model}
\end{equation}
which is formally equivalent to the Hamiltonian of a system 
of $N$ classical unit spins $\vec{s}_i$ in a random magnetic 
field $\vec{H}_i$. The variables $\theta_i\in[0,2\pi)$ 
describe the orientation of the dipole in Rochelle salt (Fig.~\ref{fig:fig1}bi), the position of layers in a liquid crystal~\cite{degennes} (Fig.~\ref{fig:fig1}bii), 
the superconducting phases of the grains~\cite{pellan} (Fig.~\ref{fig:fig1}biii), 
and the bond angles of amino acids within a protein~\cite{dill} (see Fig.~\ref{fig:fig1}biv). 
The constants $J_{ij}>0$ (usually $J_{ij}=1$) model 
the ferroelectric interactions between dipoles in piezoelectric 
quartz, liquid crystals, and protein molecules, or the Josephson 
couplings between superconducting grains. 
The adjacency matrix $A$ encodes the underlying lattice 
geometry ($A_{ij}=1$ if $i$ interacts --or is connected-- 
with $j$; $A_{ij}=0$ if not). 
Due to the positional disorder inherent in both granular 
superconductor and liquid crystals, we elect to model $A$ 
via a random regular graph with connectivity $C$. 
Vectors $\vec{H}_i=({H}_i^x,{H}_i^y)$ are random magnetic 
fields whose components $H_i^a$, $a=x,y$, are i.i.d. normal 
random variables with zero mean and variance $H_R^2$. Throughout 
the text we will restrict the discussion to Hamiltonian~\eqref{eq:model}, 
often called the random field XY-model, but in the  
Supplementary Information section~\ref{si:theory} we provide 
details on random field $O(n)$-models for general $n$ and show 
that the phenomenology of re-entrance is robust to increasing 
$n$ from the $n=2$ XY-model. 
The quantum version of the model can be obtained by adding the 
conjugate momenta in Eq.~\eqref{eq:model}, i.e. the electron 
number operators describing the effect of the charging energy 
on the superconducting grains~\cite{simanek1, efetov}. However, 
this is not the crucial ingredient underpinning the re-entrant 
phase, as we show below, and hence will not be discussed here.

It is widely believed that the principal disordering agent in 
the model described by Eq.~\eqref{eq:model} is the quenched 
disorder rather than the thermal fluctuations~\cite{imry1, fisher}. 
This belief, however, is incompatible with the phenomenon 
of re-entrance, in that there exists thermally activated 
processes that destroy the paramagnetic ground state by 
inducing a global magnetization when the system is heated 
up from zero temperature. 
Although important differences may exist in the transport 
properties, the low temperature paramagnetic phase is 
thermodynamically identical in its macroscopic properties 
(notably magnetization and susceptibility) to the higher 
temperature paramagnetic phase. In other words, the low 
temperature phase is a genuine re-entrant paramagnetic 
(or spin-fluid) phase, and not a spin-glass state~\cite{lupo}, 
as explained below.  

\begin{figure}[h]
\includegraphics[width=1\textwidth]{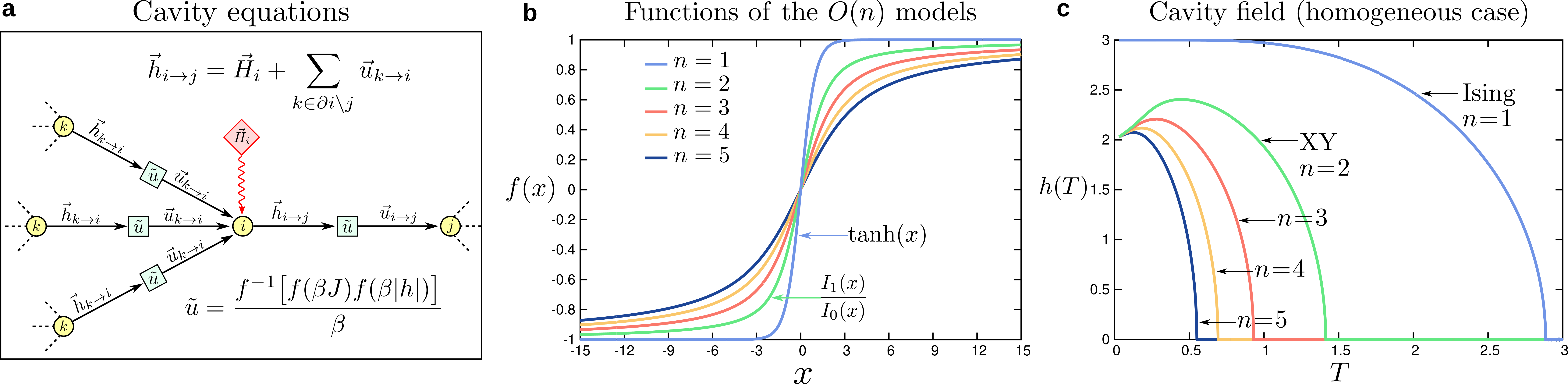} 
\caption{{\bf Definition of the model.}
\textbf{a,} Self consistent equations~\eqref{eq:messpass} 
for the order parameters (called cavity fields) of the 
$O(n)$ ferromagnetic model in a random external field on 
a random regular graph. 
Each spin $i$ receives `messages' $\vec{u}_{k\to i}$ 
containing the information about the cavity field 
$\vec{h}_{k\to i}$ and the interaction $J$ from neighboring nodes 
through the function 
$\tilde{u}(\beta, J, h)=\beta^{-1}f^{-1}[f(\beta J)f(\beta h)]$. 
Based on the messages it receives and the local random field, 
spin $i$ then broadcasts the cavity field $\vec{h}_{i\to j}$ 
to spin $j$, as prescribed by Eq.~\eqref{eq:messpass}. 
\textbf{b,} The function $f(x)$ entering in the definition of 
$\tilde{u}(\beta, J, h)$ for several values of $n$. 
\textbf{c,} Magnitude of the cavity field for the ferromagnetic model 
without random field on a random regular graph of connectivity $C=4$ 
for several values of $n$.}
\label{fig:fig2}
\end{figure}

The order parameter of the model in Eq.~\eqref{eq:model} 
is the effective field acting on spin $i$ in a modified 
graph where spin $j$ is absent, $\vec{h}_{i\to j}$, called 
{\it cavity field}~\cite{mezard} 
(see Fig.~\ref{fig:fig2}a and Supplementary Information section~\ref{si:theory}). 
The cavity fields can be thought of as `messages' exchanged 
by the spins in the graph containing the information about 
their orientation on the circle. 
Based on the information they receive, spins broadcast 
further messages, until they eventually settle in the 
directions $\theta_i$ which minimize the free-energy. 
The equations governing the flow of cavity fields in 
locally tree-like random graphs take the form  (details in 
Supplementary Information section~\ref{si:theory})
\begin{equation}
\vec{h}_{i\to j} = \vec{H}_i + \sum_{k\in\partial i\setminus j}
\tilde{u}(\beta, J_{ki}, h_{k\to i})\frac{\vec{h}_{k\to i}}{|\vec{h}_{k\to i}|}\ ,
\label{eq:messpass}
\end{equation}
where $\beta=T^{-1}$ is the inverse temperature and 
$\beta\tilde{u}(\beta, J, h) = f^{-1}[f(\beta J)f(\beta h)]$ 
with the function $f(x)$ defined as the ratio of modified 
Bessel functions $f(x) = I_1(x)/I_0(x)$, see Fig.~\ref{fig:fig2}b 
(we set henceforth $J_{ki}=J=1$). The cavity Eqs.~\eqref{eq:messpass} 
represent our first important result. 

In absence of random field, $\vec{H}_i=0$, the system 
undergoes a second order phase transition at a critical 
temperature $T_c$ defined by the condition
$\frac{J}{T_c}=f^{-1}\left(\frac{1}{C-1}\right)$,
where $\{\vec{h}_{i\to j}\neq 0\}$ (Fig.~\ref{fig:fig2}c) 
and the system magnetizes. Ferromagnetism is stable 
with respect to longitudinal fluctuations of the 
magnetization, but only marginally stable with respect 
to transverse fluctuations (see Supplementary Information section~\ref{si:theory}). 

\begin{figure}[h!]
\includegraphics[width=0.9\textwidth]{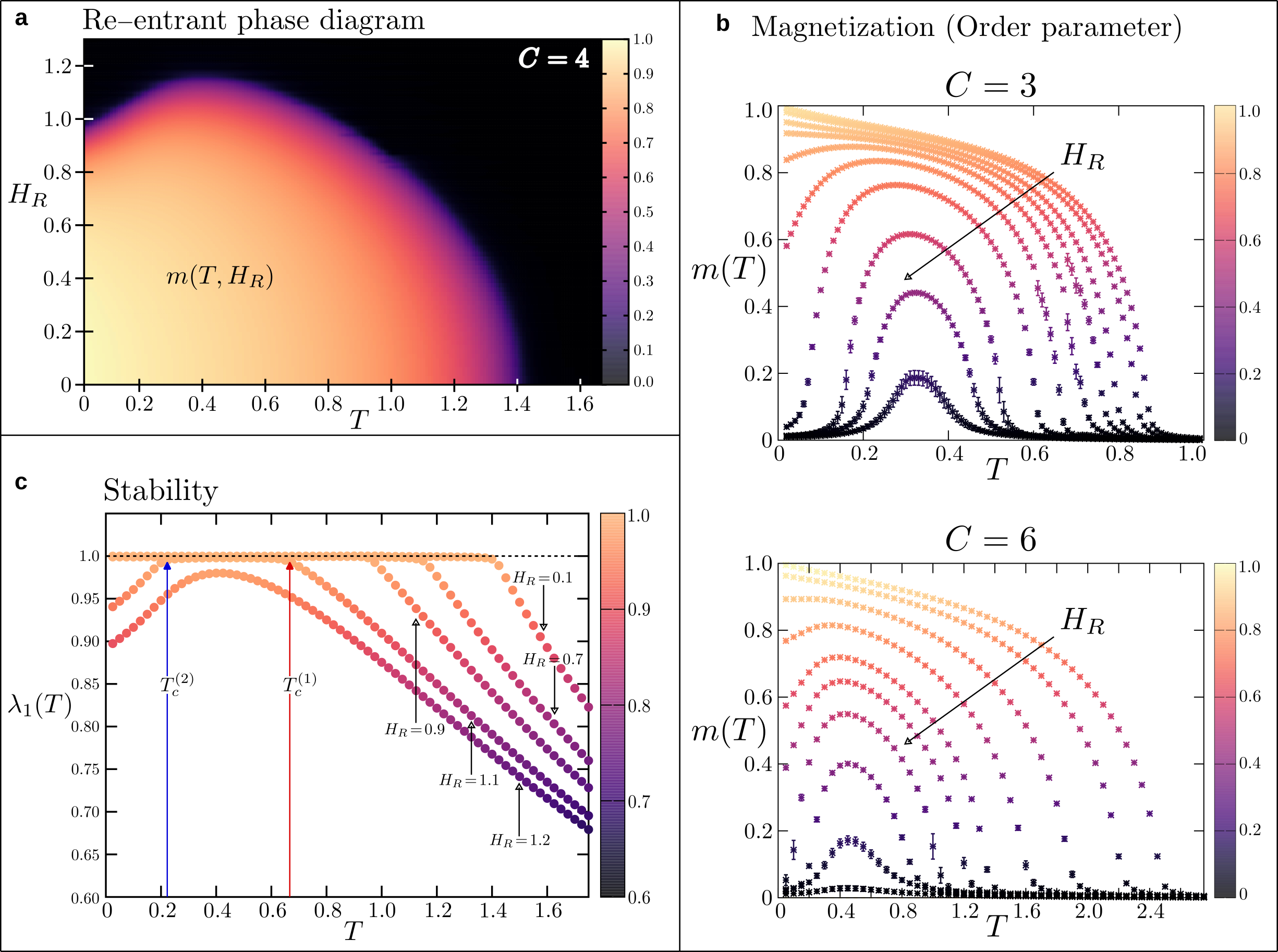} 
\caption{{\bf Re-entrant phase transitions and stability analysis.}
\textbf{a,} Phase diagram in the temperature-disorder 
$(T,H_R)$ plane of the $O(2)$ model in a Gaussian 
random field on a random regular graph with connectivity $C=4$, 
featuring a prominent re-entrant phase and non-monotonic 
behavior at low temperature.  
For $H<H_c\sim1$ the system has only one critical temperature. 
The re-entrant regime occurs for $H_c\leq H\leq H_{\rm max}\sim1.15$ 
where the system has two critical temperatures: at $T_c^{(1)}$ 
the magnetization becomes nonzero and the system orders; at 
$T_c^{(2)}<T_c^{(1)}$, the magnetization goes back to zero 
and the system re-enters into the disordered phase. 
\textbf{b,} Magnetization $m(T)$ of the $O(2)$ model in a 
Gaussian random field on random regular graphs with connectivity 
$C=3$ and $C=6$ for several values of the random field 
standard deviation $H_R$. Re-entrant phases are observed in 
both cases (error bars are s.e.m. over 30 graphs of size $N=10^6$). 
\textbf{c,} Largest eigenvalue of the stability matrix $\mathcal{M}$ 
of the $O(2)$ model in a Gaussian random field on a random regular 
graph with connectivity $C=4$ for several values of the random 
field $H_R$. The profile of $\lambda_1(T)$ are obtained by power 
iteration in Supplementary Information section~\ref{si:theory}. 
The solution is stable in the paramagnetic phases at high, 
$T>T_1^{\rm critic}$, and low, $T<T_2{\rm critic}$, temperatures, 
since $\lambda_1(T)<1$; and marginally stable in the whole 
ferromagnetic phase $T_1{\rm critic}\leq T\leq T_2{\rm critic}$ 
wherein $\lambda_1(T)=1$ (error bars are s.e.m. over 30 graphs 
of size $N=10^6$). 
}
\label{fig:fig3}
\end{figure}

The physics becomes much more interesting when we switch 
on the random field $\vec{H}_i\neq0$. Qualitatively, it seems reasonable 
that the interaction of a spin with a small random field, by 
competing with the exchange interactions, results in a 
downward shift of the critical temperature, i.e. $T_c(H_R) < T_c(0)$. 
This is precisely what we find at small random field by  
solving Eq.~\eqref{eq:messpass} on large random regular 
graphs of $N=10^6$ nodes to compute the global 
magnetization $m(H_R,T)$, shown in Fig.~\ref{fig:fig3}a. 
However, for larger values of the random field, the magnetization 
displays a dome-like profile as a function of the temperature 
(see Figure~\ref{fig:fig3}b), departing from zero at the 
critical temperature $T^{(1)}_c(H_R)$, reaching a maximum 
as the temperature decreases, and going back to zero at a 
{\bf second critical temperature} $T^{(2)}_c(H_R)$. 
Figure~\ref{fig:fig3}b shows the profile of the magnetization 
for graphs with different connectivity $C$. Remarkably, in all 
these cases we find a clear signature of a re-entrant phase 
transition into a demagnetized state at low temperatures. 
A re-entrant regime is present at any finite $C>2$ and for any 
number of components $n>1$, although the regime shrinks to zero 
with increasing $C$ and/or $n$. As such, analytically tractable 
cases such as the fully connected graph or large $n$ models do 
not exhibit re-entrance~\cite{morone23}, neither does the Ising 
model ($n=1$).
To examine whether the fixed point solution 
$\{\vec{h}^*_{i\to j}\}$ is stable we apply a small perturbation 
to the cavity fields, $\vec{h}_{i\to j}=\vec{h}^*_{i\to j} + \vec{\epsilon}_{i\to j}$, 
and expand the right-hand-side of Eq.~\eqref{eq:messpass} to first order 
in $\epsilon$, thus obtaining the linear system 
$\vec{\mathcal{E}} = \mathcal{M}\vec{\mathcal{E}}$, 
where $\vec{\mathcal{E}}$ is a vector with $2Mn$ entries 
($2M$ is the number of directed edges of the graph) obtained 
by column staking the $2M$ vectors $\vec{\epsilon}_{i\to j}$, 
and $\mathcal{M}$ is the stability matrix:  
\begin{equation}
\mathcal{M}_{i\to j, k\to l}^{\mu\nu} = 
\Big[
a(h_{k\to l})\mathcal{L}_{k\to l}^{\mu\nu} + 
b(h_{k\to l})\mathcal{T}_{k\to l}^{\mu\nu}
\Big]\mathcal{B}_{i\to j, k\to l}\ ,
\label{eq:stability_Matrix}
\end{equation}
where $\mathcal{B}_{i\to j, k\to l}$ is the non-backtracking 
matrix of the graph having non-zero entries only when 
$(k\to l, i\to j)$ form a pair of consecutive non-backtracking
directed edges, i.e. $(k\to i, i\to j)$ with $k\neq j$. 
The quantity in square bracket in Eq.~\eqref{eq:stability_Matrix} 
is the sum of the longitudinal $\mathcal{L}_{k\to l}$ 
and transverse $\mathcal{T}_{k\to l}$ projectors on the 
direction parallel and orthogonal to the cavity field 
$\vec{h}_{k\to l}$, weighted by the functions 
$a(h) = d\tilde{u}/dh$ and $b(h) = \tilde{u}/h$, respectively. 
Stability of the fixed point solution 
is controlled by the largest eigenvalue $\lambda_1(T,H_R)$ of the 
matrix $\mathcal{M}$, in that if $\lambda_1(T,H_R)<1$ a perturbation 
of the cavity fields decays to zero and the solution is stable, 
while if $\lambda_1(T,H_R)>1$ the solution is unstable. 
The word `instability' here must be understood as instability 
towards a replica symmetry broken spin-glass phase.

To familiarize with the stability matrix we first observe 
that, in absence of random field, it reduces to the tensor 
product of a $n\times n$ matrix $\mathcal{M}^{\mu\nu}$ and 
the non-backtracking matrix $\mathcal{B}$ with two distinct 
families of eigenvalues, given by $a(h)\lambda_{\mathcal{B}}$ and 
$b(h)\lambda_{\mathcal{B}}$, where $\lambda_{\mathcal{B}}$ 
is any eigenvalue of the non-backtracking matrix. 
In the paramagnetic phase ($\vec{h}=0$) we find $a(0)=b(0)=f(\beta J)$ 
and the two largest eigenvalues are degenerate and equal to  
$f(\beta J)(C-1)$, where $\lambda_{\mathcal{B}}=C-1$ is the 
largest eigenvalue of $\mathcal{B}$.  
In the ferromagnetic phase ($\vec{h}\neq0$) the degeneracy is 
lifted and we have two types of perturbations: a longitudinal 
perturbation evolving as $\vec{\epsilon}_L(t) = [(C-1)a(h)]^t \vec{\epsilon}_L(0)$; 
and a transverse one evolving as $\vec{\epsilon}_T(t) = [(C-1)b(h)]^t \vec{\epsilon}_T(0)$. 
Longitudinal perturbations eventually decay to zero, 
while transverse perturbations -- the Goldstone zero modes 
that change the orientation of the cavity field -- do not decay, 
so the solution is marginally stable along the direction 
perpendicular to the cavity field.
In Supplementary Information section~\ref{si:theory} we prove 
that the largest eigenvalue of the stability matrix is precisely 
the decay rate of the disorder-averaged connected correlation 
function. 

In presence of the random field, the study of the 
collective fluctuations becomes more complicated. 
Although we can still talk about local longitudinal 
and transverse perturbations of each cavity field 
on individual edges of the graph, this separation 
does not make sense at the global level. In fact, 
collective modes are described by the eigenvectors 
of the stability matrix, that mix all local longitudinal 
and transverse perturbations to form new hybrid collective 
modes. In practice, we are interested only in the leading 
eigen-perturbation of the stability matrix, that we call 
{\it marginal}, since it is reminiscent of the Goldstone 
mode of the pure ferromagnetic case. 
The corresponding eigenvalue $\lambda_1(T,H_R)$ is then calculated 
by Rayleigh quotient iteration (see Supplementary Information 
section~\ref{si:theory}) and shown in Fig.~\ref{fig:fig3}c. For small $H_R$, the marginal 
eigenvalue increases with decreasing temperature, reaches the 
value $\lambda_1(T,H_R)=1$ at the critical temperature $T_c^{(1)}$, 
and then stays at 1 down to zero temperature, in analogy to the pure 
case. 
However, for larger fields in the range $H_c < H_R< H_{max}$, 
we find a second critical temperature $T_c^{(2)}$ marking the 
re-entrance into the low temperature paramagnetic phase, where 
the marginal eigenvalue is strictly smaller than one (on the 
contrary, a spin-glass phase would have implied $\lambda_1(T,H_R)>1$, 
see Supplementary Information section~\ref{si:spinglass}). 
The largest eigenvalue of the stability matrix is our second and 
most important result since it contains the physical signature  
of the re-entrant phase transition and indicates that the replica 
symmetry is not broken in the re-entrant paramagnetic phase. 
Having established the existence of a re-entrant phase, we move 
to explain the physical mechanism behind it. 

We use the Jensen-Bogoliubov inequality 
to write a variational approximation $\Phi(\vec{\theta}, \mathcal{C})$ to the free-energy as $F\leq \Phi = F_0 + 
\langle H-H_0\rangle_0$, where the variational parameters 
$\vec{\theta}$ and $\mathcal{C}$ are determined by 
minimizing the approximate free energy $\Phi$. 
We find (see Supplementary Information section~\ref{si:screening}) 
that the variational free energy takes the same 
form as the original Hamiltonian~\eqref{eq:model}, 
where we simply replace the bare coupling constants 
and bare random fields with their renormalized values: 
\begin{equation}
\begin{aligned}
J_{ij}^{ren} &= J_{ij} e^{-T(\mathcal{C}_{ii}+\mathcal{C}_{jj}-2\mathcal{C}_{ij})/2}\ ,\\
H_i^{ren} &= H_i e^{-T\mathcal{C}_{ii}/2}\ ,
\end{aligned}
\label{eq:heff}
\end{equation}
while also accounting for the entropy $S\sim \log \det \mathcal{C}$ 
in the Gaussian fluctuations (see also Eq.~\eqref{eqsi:varfreeEnergy} 
in Supplementary Information section~\ref{si:screening}). 
Minimal free energy thus simply corresponds to finding the 
configuration of the angles that minimizes the effective 
energy -- just like one would do at zero temperature but 
now with renormalized coupling constants $J^{ren}$ and 
$H^{ren}$ -- while also self-consistently recomputing the 
coupling constants themselves. The latter are found by 
minimizing the free energy with respect to $\mathcal{C}$. Elementary algebra shows that this implies 
\begin{equation}
\mathcal{C}^{-1}_{ij}=\frac{\partial^2 \Phi}{\partial \theta_i\partial \theta_j}\equiv \mathcal{H}_{ij}\ ,
\label{eq:hessianT0}
\end{equation}
which simply expresses a self-consistency condition 
for the Gaussian fluctuations. 
Since the Hessian matrix $\mathcal{H}_{ij}$ is non-zero 
only on the diagonal and on the edges of the graph,  
it can be interpreted as an effective single particle 
Hamiltonian for a quantum particle hopping on a random 
regular graph with some random local energies. 
In that language, $\mathcal{C}_{ij}$ is the zero-energy 
propagator of the quantum fluctuations.

\begin{figure}[h] 
\includegraphics[width=1.0\textwidth]{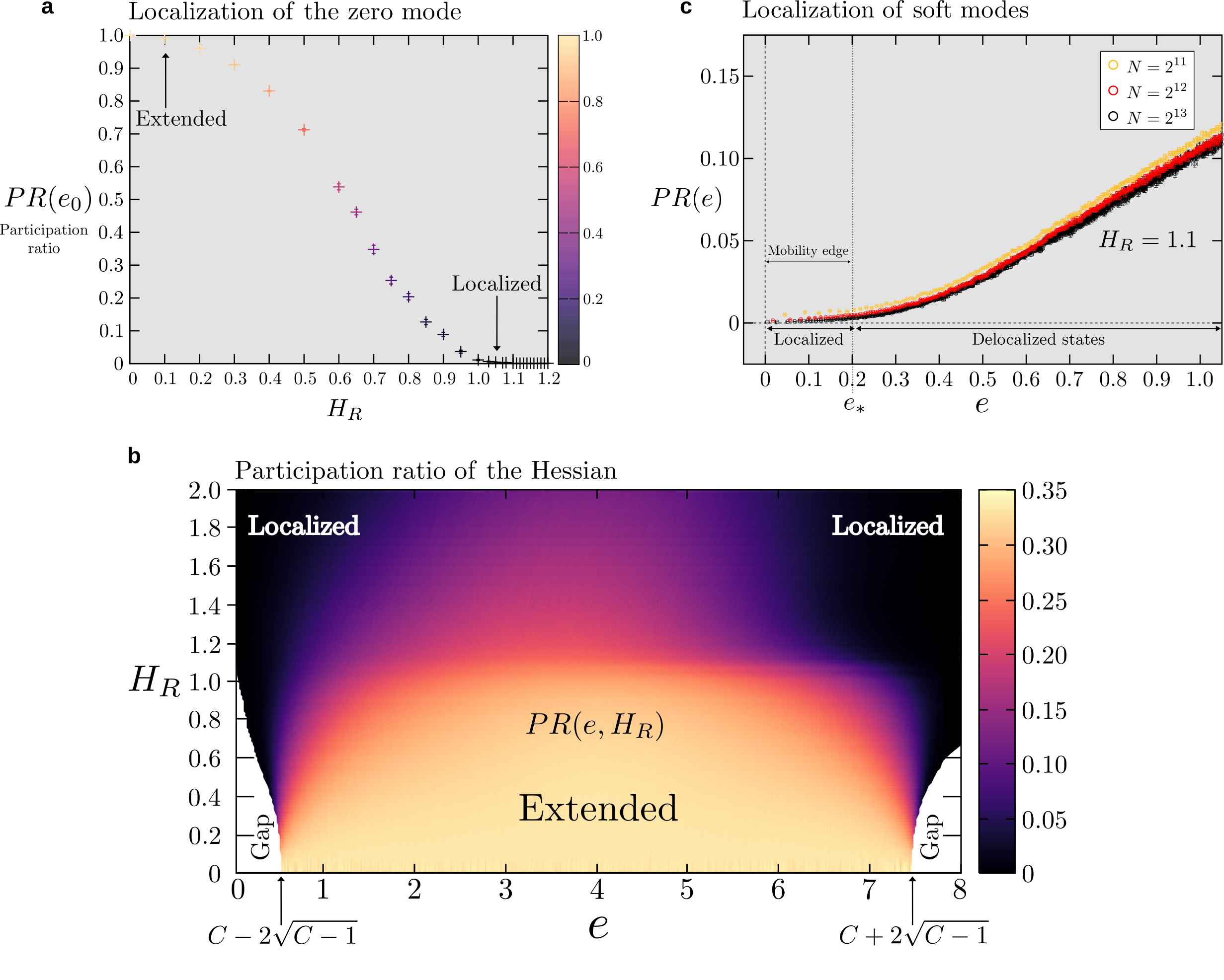} 
\caption{{\bf Localization of low-energy fluctuations.} 
\textbf{a,} 
Participation ratio of the zero-eigenmode of the Hessian 
in Eq.~\eqref{eq:hessianT0}, quantifying the degree of 
spatial localization. It is defined as 
$PR=\langle v^2\rangle^2/\langle v^4\rangle$, such that 
$PR = 1$ when the eigenmode is homogeneously delocalized 
over the entire graph and $PR = 0$ when it is localized 
on a sub-extensive number of its nodes (error bars are 
s.e.m. over 100 random regular graphs with $C=4$ and $N=2^{12}$). 
\textbf{b,} 
Participation ratio of the bulk of the Hessian's spectrum 
as a function of the eigenvalues $e$ and random field $H_R$. 
At small random field, all eigenvalues are delocalized and 
the spectrum is gapped. At $H_R\sim 1$ the spectrum becomes 
gapless. 
\textbf{c,} 
Participation ratio of the low-energy modes at $H_R=1.1$, 
showing the presence of a mobility edge such that all 
eigenmodes with eigenvalues $e\in[0,e_*]$ are fully localized. 
If the temperature is smaller than the mobility edge, 
$T\leq e_*$ only localized modes are relevant and the system thus remains paramagnetic. 
When $T$ exceeds the mobility edge, $T > e_*$, thermal 
fluctuations can excite delocalized modes and the 
system re-enters in the magnetized state. 
(Error bars are s.e.m. over 500, 300, and 100 random regular 
graphs with $C=4$ and size $N=2^{11}, 2^{12}, 2^{13}$, respectively).
}
\label{fig:figHessiandiagram}
\end{figure}

Re-entrant order is thus hidden in the finite temperature 
renormalization of the coupling constants~\eqref{eq:heff} 
and governed by the properties of the single particle 
wave functions, which have been studied extensively in 
the context of Anderson localization on random regular 
graphs~\cite{abou, aizenman11, pino, mata, tikhonov}.
Single particle states which are delocalized barely 
renormalize the effective coupling, 
$J_{ij}^{ren}\sim J_{ij}$, but the random fields 
get screened out, $H^{ren}_i\ll H_i$, since those 
eigenstates would have 
$\mathcal{C}_{ii}\sim \mathcal{C}_{jj}\sim \mathcal{C}_{ij}$. 
Conversely, states that are very well localized screen 
out the couplings more than the random fields, since 
they have vanishing correlations $\mathcal{C}_{ij}\sim 0$. 
To understand which effect is the strongest at low 
temperature, one has to understand the zero-temperature 
structure of the Hessian.
For small random field the Hessian is just the graph 
Laplacian with diagonal disorder. The latter has a 
ground state gapped from the rest of the spectrum~\cite{clark18}, 
and it has been rigorously shown that all the states 
are extended~\cite{aizenman11} below a critical disorder, 
as seen in Figs.~\ref{fig:figHessiandiagram}a,b. 
As a consequence the random fields get screened 
out more than the couplings, showing that the finite 
temperature ferromagnet is stable to weak disorder. 
Upon increasing the disorder, the ground 
state localizes, shown in Fig.~\ref{fig:figHessiandiagram}a, 
the gap between the ground state and the bulk states closes, 
shown in Fig.~\ref{fig:figHessiandiagram}b, 
and long range order is lost. 
At that point, the system remains gapless and a 
mobility edge forms with localized low energy states~\cite{nowak, igarashi}, 
while the bulk is still extended, as seen in Fig.~\ref{fig:figHessiandiagram}c. 
It is in this regime that the system is sufficiently 
strongly correlated to display re-entrant order, 
in that, when $T$ exceeds the mobility edge, thermal 
agitation can excite delocalized modes which synchronize 
the fluctuations of the order parameter and, in doing so, 
magnetize the system for $H_R \in [Hc, H_{max}]$. 

Taken together, our results unveil the microscopic 
molecular mechanism behind re-entrant phase transitions 
and elucidate the role of localization of collective 
soft-modes in the process. 

\bigskip
\bigskip
\bigskip

\noindent
{\bf \large Data availability}
Data that support the findings of this study 
can be generated by solving the cavity equations, 
computing the largest eigenvalue of the stability 
matrix, and diagonalizing the zero-temperature hessian 
of the model given in equation~\eqref{eq:model}.

\medskip

\noindent
{\bf \large Code availability}
The source code to solve the cavity equations, 
compute the largest eigenvalue of the stability 
matrix, and the zero-temperature hessian are 
available upon request. 

\medskip

\noindent
{\bf \large Acknowledgments} 
The  Flatiron  Institute  is  a  division  of  the  
Simons Foundation. We acknowledge support from Air 
Force Office of Scientific Research(AFOSR): Grant 
FA9550-21-1-0236. 

\medskip

\noindent
{\bf \large Author contributions}
FM and DS conceived the study, performed all the analytic 
calculations, implemented the code, and wrote the manuscript.  

\medskip

\noindent
{\bf \large Additional information}
Supplementary Methods accompany this paper. 

\medskip

\noindent
{\bf \large Competing interests} 
All authors declare no competing interests. 

\medskip

\noindent
{\bf \large Correspondence} should be addressed to FM at: fm2452@nyu.edu

\newpage

\onecolumngrid

\newpage

{\centering

{\normalsize \bf Supplementary Information: Re-entrant phase transitions induced by localization of zero-modes}

\vspace{0.5cm}

Flaviano Morone$^{1}$, Dries Sels$^{1, 2}$\\

\vspace{0.1in}

{\it
$^1$Department of Physics, New York University, New York, New York 10003, USA\\
$^2$Center for Computational Quantum Physics, Flatiron Institute,\\
162 Fifth Avenue, New York, New York 10010 USA\\
}}

\vspace{1cm}

\onecolumngrid

\renewcommand{\thesection}{S\arabic{section}}  
\renewcommand{\thefigure}{S\arabic{figure}}

\setcounter{equation}{0}
\setcounter{figure}{0}

\tableofcontents

\clearpage

\section{The random field $O(n)$ model on random graphs}
\label{si:theory}

The random field $O(n)$ model on random graphs 
describes a great variety of important physical 
systems while being analytically tractable 
and phenomenologically different from the, perhaps 
most popular random field Ising model. The main 
difference is the existence of a remarkable re-entrant 
phase transition with a rich physical content that 
is the leitmotif of the present paper. 
From the mathematical standpoint, our main result 
is the discovery of a closure scheme to approximtely 
solve the cavity equations and thus compute the local magnetizations efficiently on graphs with millions 
of nodes.
Furthermore, by perturbing the fixed point solution 
to the cavity equations, we derive the analytical 
form of the stability matrix, which in turn allows 
us to compute the susceptibility from the largest 
eigenvalue of said matrix. By analyzing the condition 
for the stability of the fixed point solution we draw 
the full phase diagram of the model in the temperature-random field plane and, in doing that, we discover a re-entrant disordered phase at low temperature in a range of values 
of the random-field strength. 
Finally, to unlock our physical understanding of the 
re-entrance, we study the spectrum of the low temperature excitations and we conclude that the re-entrant phase 
transition occurs as a consequence of the spatial 
localization of soft-modes on a sub-extensive number 
of sites of the random graph. 

We start with the derivation of the closure scheme 
for the cavity equations, discussed next. 

\subsection{Cavity equations}
In this section we derive the cavity equations for the 
Hamiltonian 
\begin{equation}
\mathcal{H} = -\frac{J}{2}\sum_{i,j=1}^N A_{ij}\vec{s_i}\cdot\vec{s_j} - 
\sum_{i=1}^N\vec{H}_i\vec{s_i}\ ,
\label{eq:OnHamiltonian}
\end{equation}
where $J>0$ is the ferromagnetic interaction strength, 
$A_{ij}$ is the adjacency matrix of the random graph, 
$\vec{s}_i$ are $n$-dimensional unit spins, $|\vec{s}_i|=1$, 
and $\vec{H}_i$ is a local magnetic field whose $n$ 
components are i.i.d. normal random variables with 
zero mean and variance $H_R^2$. 
To write down the cavity equations in the simple form 
given in Eq.~\eqref{eq:messpass} in the main text we 
need two ingredients. The first one is the following 
integral
\begin{equation}
G_0(\vec{a}) = \int_{\mathcal{S}_{n-1}}d\vec{s}\ e^{\vec{s}\cdot\vec{a}}\ ,
\label{eq:g0}
\end{equation}
where $\mathcal{S}_{n-1}$ is the unit $(n-1)$-sphere 
defined as $\mathcal{S}_{n-1} = 
\{\vec{s}\in\mathbb{R}^{n} : ||s|| = 1\}$
Notice that $G_0(\vec{a})$ depends only on 
the magnitude of the vector $\vec{a}$. To see 
this, let us consider a rotation $R$ and evaluate 
$G_0(R\vec{a})$:
\begin{equation}
G_0(R\vec{a}) = \int_{\mathcal{S}_{n-1}}d\vec{s}\ 
\exp\left(\sum_{ij}s_iR_{ij}a_j\right)\ .
\end{equation}
By making a change of variables $s_j' = \sum_iR_{ij}s_i$ 
and observing that the integration measure is invariant, $d\vec{s}^{\ '}=d\vec{s}$ (since $R$ is an isometry), we 
conclude that
\begin{equation}
G_0(R\vec{a}) = G_0(\vec{a})\ \to\ 
G_0(\vec{a}) = G_0(|a|)\ .
\end{equation}
Therefore, without loss of generality, we can 
choose $\vec{a}=(|a|,0,...,0)$, thus finding 
\begin{equation}
\begin{aligned}
G_0(|a|) &= \int_{\mathcal{S}_{n-1}}d\vec{s}\ e^{|a|s_1} = 
\int_{-1}^{1}ds_1\ e^{|a|s_1}\int_{-1}^{1}ds_2...ds_n\ 
\delta\left(\sqrt{s_1^2+...+s_n^2}-1\right) = \\
&= 
\int_{-1}^{1}ds_1\ e^{|a|s_1}\int_{0}^{1}r^{n-2}dr\ 
\delta\left(\sqrt{s_1^2+r^2}-1\right)\int d\Omega_{n-2} = \\
&=
\frac{2\big[\pi^{(n-1)/2}\big]}{\Gamma\Big(\frac{n-1}{2}\Big)} 
\int_{-1}^{1}ds_1\ e^{|a|s_1}\Big(1-s_1^2\Big)^{(n-3)/2}\ .
\end{aligned}
\label{eq:g0}
\end{equation}
The second ingredient is the following integral
\begin{equation}
\vec{G}(\vec{a}) = 
\int_{\mathcal{S}_{n-1}}d\vec{s}\ \vec{s}\ e^{\vec{s}\cdot\vec{a}}\ .
\end{equation}
By applying a rotation $R$ to $\vec{a}$ we 
find that
\begin{equation}
\vec{G}(R\vec{a}) = R\vec{G}(\vec{a})\ ,
\end{equation}
hence the most general form of $\vec{G}(\vec{a})$ is 
\begin{equation}
\vec{G}(\vec{a}) = G_1(|a|)\hat{a}\ ,
\label{eq:G1}
\end{equation}
where $\hat{a}$ is a unit vector in the direction of $\vec{a}$, 
and $G_1(|a|)$ is given by 
\begin{equation}
G_1(|a|) = 
\int_{\mathcal{S}_{n-1}}d\vec{s}\ \vec{s}\cdot\hat{a}\ e^{\vec{s}\cdot\vec{a}}\ .
\end{equation}
To evaluate $G_1(|a|)$ we can choose, again 
without loss of generality, $\vec{a}=(|a|,0,...,0)$, 
thus obtaining
\begin{equation}
G_1(|a|) = 
\frac{2\big[\pi^{(n-1)/2}\big]}{\Gamma\Big(\frac{n-1}{2}\Big)} 
\int_{-1}^{1}ds_1\ e^{|a|s_1}\ s_1\ \Big(1-s_1^2\Big)^{(n-3)/2}\ .
\label{eq:g1}
\end{equation}
Next we write down the self-consistent equations for the 
cavity marginals $p_{i\to j}(\vec{s}_i)$ of the model in Eq.~\eqref{eq:OnHamiltonian} that read
\begin{equation}
p_{i\to j}(\vec{s}_i) \approxeq e^{\beta \vec{H}_i\cdot\vec{s}_i}
\prod_{k\in\partial i\setminus j}\int_{\mathcal{S}_{n-1}}d\vec{s}_k\ 
e^{\beta J \vec{s}_i\cdot\vec{s}_k}\ p_{k\to i}(\vec{s}_k)\ ,
\label{eq:cavityEq}
\end{equation}
where `$ \approxeq$' means `equal up to a normalization factor'~\cite{montanari}. 
The function $p_{i\to j}(\vec{s}_i)$ can always be 
written as 
\begin{equation}
p_{i\to j}(\vec{s}_i)\approxeq e^{\beta W_{i\to j}(\vec{s}_i)}\ ,
\end{equation}
and the function $W_{i\to j}(\vec{s}_i)$ can be 
parametrized as
\begin{equation}
W_{i\to j}(\vec{s}_i) = \sum_{a=1}^n h^a_{i\to j}s_i^a + 
\sum_{a,b=1}^n h^{ab}_{i\to j}s_i^as_i^b + ... + 
\sum_{a_1,...,a_\nu=1}^n h^{a_1...a_\nu}_{i\to j}s_i^{a_1}...s_i^{a_\nu} +...
\label{eq:Wfunc}
\end{equation}
where in addition to a cavity vectorial field, $h^{a}_{i\to j}$, 
we have included a second-rank matrix $h^{ab}_{i\to j}$ 
and a general $\nu^{\rm th}$-rank tensor $h^{a_1...a_\nu}_{i\to j}$. To make progress, however, we retain in the expansion 
of the function $W_{i\to j}(\vec{s}_i)$ only the vectorial 
term, i.e., we parametrize the cavity marginal as 
\begin{equation}
p_{i\to j}(\vec{s}_i)\approxeq e^{\beta \vec{h}_{i\to j}\cdot\vec{s}_i}\ ,
\label{eq:pij}
\end{equation}
thus using what we may call the dipolar (or vectorial) approximation~\cite{javanmard}. 
This approximation amounts to neglect the quadrupolar terms 
$s_i^as_i^b$ and higher order multipolar contributions, 
which may, in principle, be included perturbatively once a 
solution at the leading dipolar order has been obtained, 
that is what we work out next. 
Plugging Eq.~\eqref{eq:pij} into Eq.~\eqref{eq:cavityEq} 
we obtain 
\begin{equation}
e^{\beta h_{i\to j}\cdot\vec{s}_i} \approxeq 
e^{\beta \vec{H}_i\cdot\vec{s}_i}
\prod_{k\in\partial i\setminus j}\int_{\mathcal{S}_{n-1}}d\vec{s}_k\ 
e^{\beta J \vec{s}_i\cdot\vec{s}_k + \beta h_{k\to i}\cdot\vec{s}_k}\ .
\label{eq:cavityEq2}
\end{equation}
The goal here is to find a closed self-consistent 
equation for the set of cavity fields $\{\vec{h}_{i\to j}\}$.
To this end, we search for a solution of the integral on 
the r.h.s of Eq.~\eqref{eq:cavityEq2} of the form 
\begin{equation}
\int_{\mathcal{S}_{n-1}}d\vec{s}\ 
e^{\beta J \vec{r}\cdot\vec{s} + \beta \vec{h}\cdot\vec{s}} = 
A(\beta J, \beta |h|, \beta |u|)\ e^{\beta \vec{u}\cdot\vec{r}}\ .
\label{eq:integral_u}
\end{equation}
To find $A(\beta J,\beta |h|,\beta |u|)$ we integrate 
over $d\vec{r}$ on both sides of Eq.~\eqref{eq:integral_u} 
and then use Eq.~\eqref{eq:g0} to get 
\begin{equation}
A(\beta J,\beta |h|,\beta |u|) = 
\frac{G_0(\beta J)G_0(\beta |h|)}{G_0(\beta |u|)}\ ,
\end{equation}
where the vector $\vec{u}$ is usually called cavity bias. 
To find $\vec{u}$ we multiply by $\vec{r}$ both sides 
of Eq.~\eqref{eq:integral_u}, integrate over $\vec{r}$ 
and use Eq.~\eqref{eq:G1} to obtain 
\begin{equation}
G_1(\beta J)G_1(\beta |h|)\hat{h} = A(\beta J,\beta |h|,\beta |u|)
G_1(\beta |u|)\hat{u}\ .
\end{equation}
We deduce that $\vec{u}$ is a vector in the same 
direction of $\vec{h}$ whose magnitude is given by
\begin{equation}
|u| \equiv \tilde{u}(h) = \frac{1}{\beta}
f^{-1}[f(\beta J)f(\beta |h|)]\ ,
\end{equation}
where the function $f(x)$ is defined by
\begin{equation}
f(x) = \frac{G_1(x)}{G_0(x)} = 
\frac{\int_{-1}^{1}dy\ y\left(1-y^2\right)^{(n-3)/2}e^{xy}}
{\int_{-1}^{1}dy\ \left(1-y^2\right)^{(n-3)/2}e^{xy}} = 
\frac{d}{dx}\log \int_{-1}^{1}dy\ \left(1-y^2\right)^{(n-3)/2}e^{xy}\ ,
\label{eq:fx}
\end{equation}
and we have dropped the dependence of the function 
$\tilde{u}(h)$ from $\beta$ and $J$ to lighten the 
notation.
For $n=1,2,3,4,5$ the function $f(x)$ is shown in Fig.~\ref{fig:fig2}b 
and explicitly given by the following expression
\begin{equation}
f(x) = \left\{
\begin{matrix}
\tanh(x) & n = 1\ ,\\
\frac{I_1(x)}{I_0(x)} & n = 2\ ,\\
\coth(x) -\frac{1}{x} & n = 3\ ,\\
\frac{1}{2}\frac{I_1(x) - I_3(x)}{I_0(x) - I_2(x)} & n = 4\ ,\\
 \frac{(x^2 + 3)\tanh(x) - 3x}{x^2 - x\tanh(x)} & n = 5\ ,
\end{matrix}
\right.
\end{equation}
where $I_k(x)$ is the modified Bessel function of the 
first kind of order $k$~\cite{arfken}. Notice that the 
function $f(x)=\coth(x)-1/x$ for $n=3$ is the well known 
Langevin function often encountered in the classical 
theory of magnetism. 

Using the previous results we can turn the self-consistent 
equations for the cavity marginals in Eq.~\eqref{eq:cavityEq} 
into self-consistent equations for the cavity fields in the 
form given in Eq.~\eqref{eq:messpass} in the main text, 
that we rewrite below
\begin{equation}
\begin{aligned}
\vec{h}_{i\to j} &= \vec{H}_i + \sum_{k\in\partial i\setminus j}\vec{u}_{k\to i}\ ,\\
\vec{u}_{k\to i} &= \tilde{u}(|h_{k\to i}|)\hat{h}_{k\to i}\ ,
\end{aligned}
\label{eq:cavityEqs}
\end{equation}  
where $\hat{h}_{k\to i}$ is a unit vector in the direction 
of the cavity field, $\hat{h}_{k\to i}=\vec{h}_{k\to i}/|\vec{h}_{k\to i}|$. 
We conclude this section by noticing that Eqs.~\eqref{eq:cavityEqs} can be interpreted 
as distributional equations for the probability 
distributions $P(\vec{h})$ and $Q(\vec{u})$ 
which satisfy the self-consistent equations 
\begin{equation}
\begin{aligned}
P(\vec{h}) &= \mathbb{E}_{H}\int \Bigg[\prod_{k=1}^{C-1}d\vec{u}_k\ Q(\vec{u}_k)\Bigg]
\prod_{\alpha}\delta\Bigg[h^{\alpha} - H^{\alpha} - 
\sum_{k=1}^{C-1}u_k^{\alpha}\Bigg]\ ,\\
Q(\vec{u}) &= \int d\vec{h}\ P(\vec{h})\ 
\prod_{\alpha}\delta\Bigg[u^{\alpha} - \tilde{u}(\beta, J, |\vec{h}|)
\frac{h_{\alpha}}{|\vec{h}|}\Bigg]\ . 
\end{aligned}
\end{equation}

\subsubsection{Zero temperature cavity equations}
\label{sec:zeroTlimit}
It is interesting to derive the zero temperature limit 
of the cavity equations for a general $O(n)$ model, 
in that the final result displays a very weak dependence 
on $n$. 

To achieve this goal, it is sufficient to compute the 
asymptotic behavior of the function $f(x)$ at large 
argument at order $O(x^{-1})$. 
To start let us consider the function $g_0(x)$ defined 
by the integral 
\begin{equation}
g_0(x) = \int_{-1}^{1}dy\ \left(1-y^2\right)^{(n-3)/2}e^{xy}\ ,
\end{equation}
which we recognize as the denominator in the definition of the 
function $f(x)$. 
After a change of variables $y=\cos(\theta)$, and setting 
$n-2=m$, we can write $g_0(x)$ as
\begin{equation}
g_0(x) = \int_{0}^{\pi}d\theta\ \sin(\theta)^m\ e^{x\cos(\theta)}\ .
\end{equation}
We multiply both sides by $\sqrt{x}e^{-x}$
\begin{equation}
\sqrt{x}e^{-x}g_0(x) = \sqrt{x}\int_{0}^{\pi}d\theta\ \sin(\theta)^m\ e^{-x[1-\cos(\theta)]}\ ,
\end{equation}
and make the change of variables $\phi = \sqrt{x}\theta$, 
thus obtaining
\begin{equation}
\begin{aligned}
\sqrt{x}e^{-x}g_0(x) &= 
\int_{0}^{\pi\sqrt{x}}d\phi\ 
\left(\sin\frac{\phi}{\sqrt{x}}\right)^m
\exp\left[-x\left(1-\cos\frac{\phi}{\sqrt{x}}\right)\right] =\\
&\sim 
\int_{0}^{\infty}d\phi\ 
\left(\frac{\phi}{\sqrt{x}}\right)^m 
\exp\left(-\frac{\phi^2}{2}\right) = 
\frac{(\sqrt{2})^{m-1}}{x^{m/2}}\Gamma\left(\frac{m+1}{2}\right)\ .
\end{aligned}
\label{eq:g0asympt}
\end{equation}
Next we consider the function $g_1(x)$
\begin{equation}
g_1(x) = \int_{-1}^{1}dy\ y\left(1-y^2\right)^{(n-3)/2}e^{xy}\ ,
\end{equation}
which is the numerator in the definition of $f(x)$, and 
can be rewritten as 
\begin{equation}
g_1(x) = \int_{0}^{\pi}d\theta\ \cos(\theta)\sin(\theta)^m\ 
e^{x\cos(\theta)}\ .
\end{equation}
Using the same manipulations leading to Eq.~\eqref{eq:g0asympt} 
we find the following asymptotic behavior of $g_1(x)$
\begin{equation}
\sqrt{x}e^{-x}g_1(x) \sim 
\frac{(\sqrt{2})^{m-1}}{x^{m/2}}
\Gamma\left(\frac{m+1}{2}\right)
\left[1-\frac{(m+1)}{2}\frac{(m+3)}{3}\frac{1}{x}\right]\ .
\label{eq:g1asympt}
\end{equation}
Taking the ratio of $g_1(x)$ and $g_0(x)$ and substituting 
$m=n-2$ we find the following asymptotic behavior of $f(x)$ 
at large $x$
\begin{equation}
f(x) = 1 - \frac{n^2-1}{6x} + O(x^{-2})\ \ \ {\rm for}\ \ \ x\to \infty\ .
\label{eq:fasympt}
\end{equation}
We note that this expansion is valid for $n>1$. 
The case $n=1$, corresponding to the Ising model, needs to be 
treated in a different way. Anyway, it is easy to show that 
\begin{equation}
f(x)\sim 1 - 2e^{-2x}\ \ \ {\rm for}\ \ \ n = 1\ .
\end{equation}
As a consequence the $O(n)$ model for $n\geq 2$ is fundamentally 
different from the Ising model, in that the function $f(x)$ 
converges to $1$ algebraically in the $O(n)$ model, and hence 
much more slowly than in the Ising model, where the convergence 
is exponentially fast. This result leads also to a very different 
form of the zero temperature cavity equations, as explained 
next. 

To compute the zero temperature limit we need one more 
ingredient, i.e., the inverse of the function $f(x)$. 
It is easy to check that, for large $x$, $f^{-1}(x)$ 
must have the following form 
\begin{equation}
f^{-1}(x) \sim \frac{n^2-1}{6(1-x)}\ \ \ {\rm for}\ \ \ x\to \infty\ ,
\end{equation}
which indeed satisfies the condition $f^{-1}(f(x))=x$. 
At this point we have all that we need to compute the 
limit for $\beta\to\infty$ of the function $\tilde{u}(h)$ 
which is 
\begin{equation}
\lim_{\beta\to\infty}\tilde{u}(h) = 
\lim_{\beta\to\infty}\frac{1}{\beta}
f^{-1}[f(\beta J)f(\beta |h|)] = 
\frac{J|h|}{J+|h|}\ .
\end{equation}
Remarkably, this limit does not depend on $n$; hence the 
function $\tilde{u}(h)$ at zero temperature becomes universal 
for $n>1$. Knowledge of the function $\tilde{u}(h)$ allows 
us to write down the zero temperature cavity equations, which 
read
\begin{equation}
\boxed{
\vec{h}_{i\to j} = \vec{H}_i + 
\sum_{k\in\partial i\setminus j}
\frac{J}{J + |\vec{h}_{k\to i}|}\ \vec{h}_{k\to i}
}\ .
\label{eq:zeroTcavEq}
\end{equation}

\subsection{Observables}
\subsubsection{Magnetization}
\begin{figure}[h!]
\includegraphics[width=0.88\textwidth]{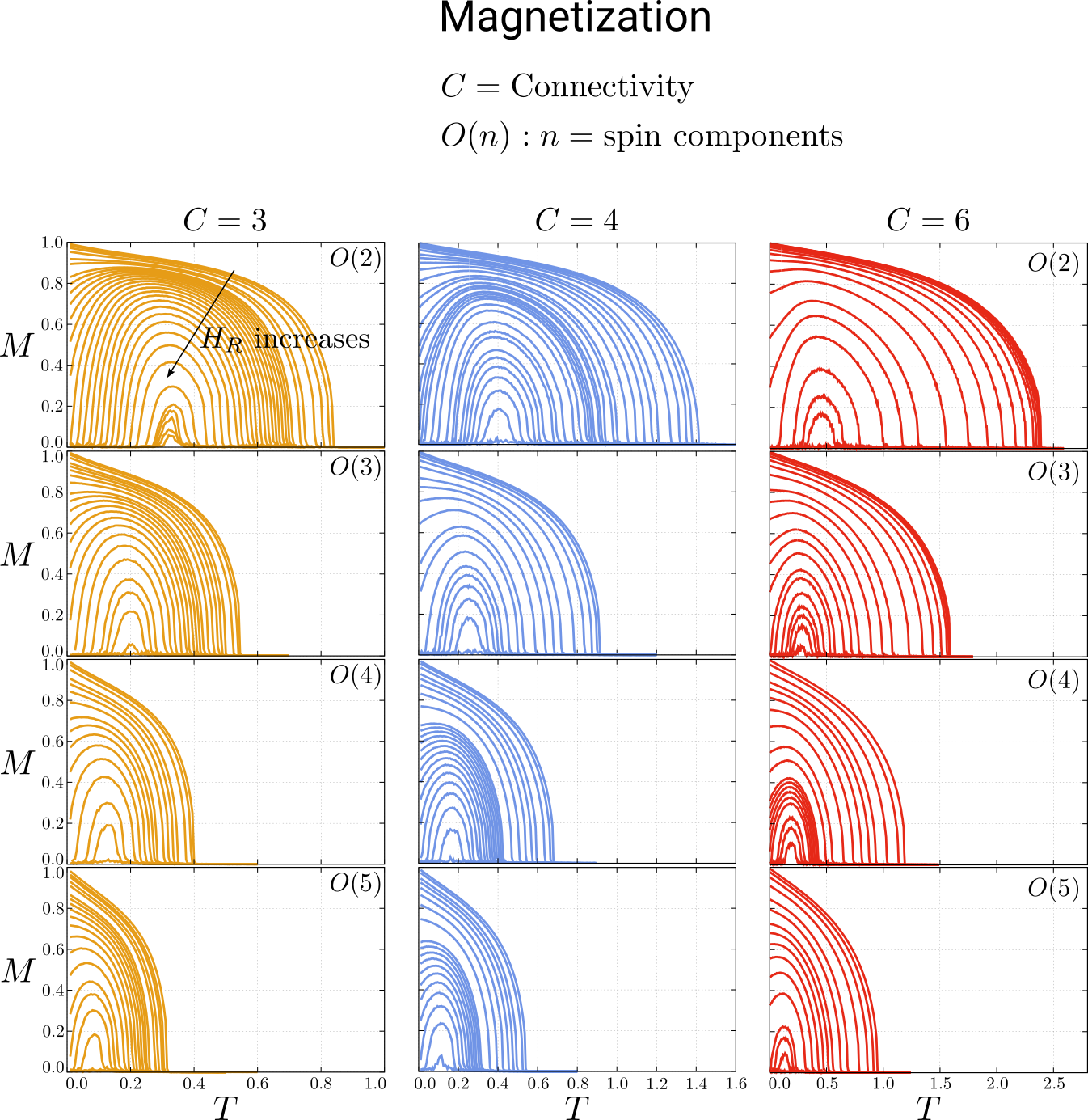} 
\caption{ Magnetization of the random field $O(n)$ 
model on random regular graphs with $N=10^6$ nodes 
and connectivity $C=3,4,6$ for different values of 
the number of spin degrees of freedom $n=2,3,4,5$. 
}
\label{fig:magSI}
\end{figure}

The solution to the cavity equations~\eqref{eq:cavityEqs} 
allows us to compute all the relevant observables and 
thermodynamic quantities. 
In particular, we can compute the single spin marginal as 
\begin{equation}
p_i(\vec{s}_i) = 
\frac{e^{\beta \vec{H}_i\cdot\vec{s}_i}}{Z_i}
\prod_{k\in\partial i}\int_{\mathcal{S}_{n-1}}d\vec{s}_k\ 
e^{\beta J \vec{s}_i\cdot\vec{s}_k}\ p_{k\to i}(\vec{s}_k)\ ,
\label{eq:1spinmarg}
\end{equation}
and from it the local magnetization as 
\begin{equation}
\vec{m}_i = \int_{\mathcal{S}_{n-1}}d\vec{s}_i\ p_i(\vec{s}_i)\ \vec{s}_i = 
f(\beta|\vec{h}_i|)\ \hat{h}_i\ ,
\label{eq:localmag}
\end{equation}
where $\vec{h}_i$ is the total magnetic field acting 
on site $i$ containing the contributions from the random 
field and the cavity biases sent to $i$ from the neighboring 
spins $k\in\partial i$, given by 
\begin{equation}
\vec{h}_i = \vec{H}_i + \sum_{k\in\partial i}\vec{u}_{k\to i}\ .
\label{eq:cavity}
\end{equation}
The total magnetization $\vec{M}$ is defined as 
\begin{equation}
\vec{M} = \frac{1}{N}\sum_i \vec{m}_i\ , 
\label{eq:totalmag}
\end{equation}
and its magnitude by $M = \Big(\sum_{a=1}^n M_a^2\Big)^{1/2}$. 
In Fig.~\ref{fig:magSI} we plot $M$ as a function of $T$ for 
random graphs of different connectivity $C=3,4,6$ and for 
different values of the spin components $n=2,3,4,5$. 
To compute $M$ we first solve Eq.~\eqref{eq:cavityEqs} 
on a given random regular graph with $N=10^6$ nodes, 
then we compute the local fields $\vec{h}_i$ 
and from them the local magnetization using Eq.~\eqref{eq:localmag}. 
Finally we compute the total magnetization 
and its magnitude using Eq.~\eqref{eq:totalmag}.  

\subsubsection{Free energy}
\begin{figure}[h!]
\includegraphics[width=0.95\textwidth]{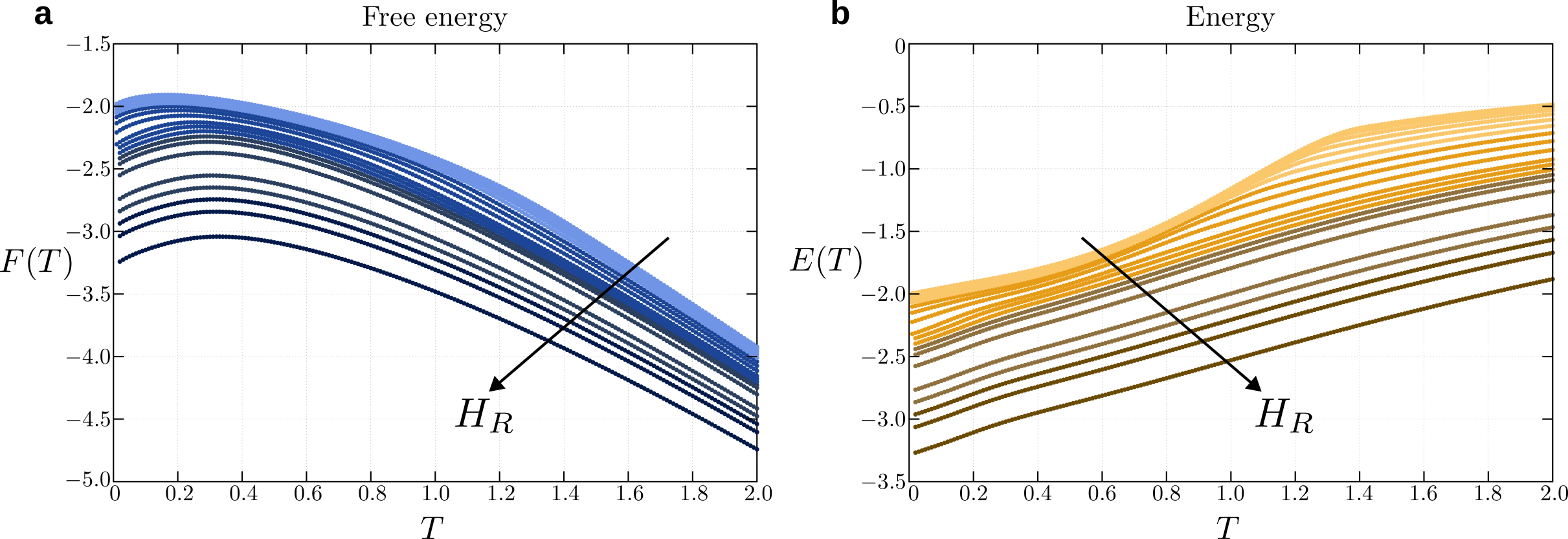} 
\caption{Free energy ({\bf a}) and internal energy ({\bf b}) 
of the RF $O(2)$ model on a random regular graph of connectivity 
$C=4$ and size $N=2\times 10^6$ for several values of the random 
field strength $H_R$. 
}
\label{fig:observableSI}
\end{figure}

Similarly to the single spin marginal, we can compute the 
joint distribution of two spins 
$\vec{s}_i$ and $\vec{s}_j$ sharing an edge as 
\begin{equation}
p_{ij}(\vec{s}_i, \vec{s}_j) = \frac{1}{Z_{ij}}
e^{\beta J\vec{s}_i\cdot \vec{s}_j}
p_{i\to j}(\vec{s}_i)p_{j\to i}(\vec{s}_j)\ .
\label{eq:2spinmarg}
\end{equation}
Of particular importance are the normalization factors $Z_{i}$ 
in Eq.~\eqref{eq:1spinmarg} and $Z_{ij}$ in Eq.~\eqref{eq:2spinmarg}, 
the knowledge of which allows us to compute the free energy $F$ of 
the model~\cite{montanari}. A simple calculation gives 
\begin{equation}
-\beta F = \frac{CN}{2}\log G_0(\beta J) + 
\left(1-\frac{C}{2}\right)\sum_{i=1}^N \log G_0(\beta |h_i|) + 
\frac{1}{2}\sum_{i=1}^N \sum_{j\in\partial i}
\log\frac{G_0(\beta |h_{j\to i}|)}{G_0(\beta |u_{j\to i}|)}\ ,
\label{eq:FreeEnergy}
\end{equation}
where $G_0(x)$ is the function defined in Eq.~\eqref{eq:g0}. 
The free energy is shown in Fig.~\ref{fig:observableSI}a for 
the case of the XY model ($n=2$) on a given RRG with connectivity 
$C=4$ and size $N=2\times 10^6$. 

The same one- and two-spins marginals given by 
Eqs.~\eqref{eq:1spinmarg} and~\eqref{eq:2spinmarg} allow us 
to compute the internal energy $U$ as 
\begin{equation}
U = -\sum_{\langle ij\rangle}J
\int d\vec{s}_id\vec{s}_j\ \vec{s}_i\cdot \vec{s}_j\ p_{ij}(\vec{s}_i, \vec{s}_j) - \sum_i\vec{H}_i\cdot \int d\vec{s}_i\ 
\vec{s}_i\ p_i(\vec{s}_i)\ .
\end{equation}
However, since the free energy in Eq.~\eqref{eq:FreeEnergy} 
is variational~\cite{montanari}, the energy can be obtained 
more easily by computing the explicit derivative with respect 
to $\beta$ without deriving with respect to $\vec{h}_{i\to j}$ 
and thus we find 
\begin{equation}
\begin{aligned}
-U = -\frac{\partial \beta F}{\partial\beta} &= 
\frac{CN}{2}Jf(\beta J) + 
\left(1-\frac{C}{2}\right)\sum_{i=1}^N |h_i| f(\beta |h_i|)\ + 
\frac{1}{2}\sum_{i=1}^N \sum_{j\in\partial i}
\Big[|h_{j\to i}|f(\beta |h_{j\to i}|) - |u_{j\to i}|f(\beta |u_{j\to i}|)\Big] \ ,
\end{aligned}
\label{eq:Energy}
\end{equation}
which is shown in Fig.~\ref{fig:observableSI}b for the case of 
the XY model ($n=2$) on a given RRG with connectivity $C=4$ and 
size $N=2\times 10^6$. 
Eventually, knowing $F$ and $U$, we can compute the entropy $S$ 
as $S=\beta(U-F)$. 

\subsection{Stability analysis}
The complete analysis of the model requires the study of the 
stability of the fixed point solution $\{\vec{h}^*_{i\to j}\}$. 
To analyze the linear stability we apply a small perturbation 
to the fixed point cavity fields as $\vec{h}_{i\to j}=\vec{h}^*_{i\to j} + \vec{\epsilon}_{i\to j}$, plug it into Eq.~\eqref{eq:cavityEqs} 
and expand the r.h.s. to first order in $\epsilon$, thus obtaining 
the following system of linear equations for the perturbations
\begin{equation}
\vec{\epsilon}_{i\to j} = \sum_{k\in\partial i\setminus j}
a(h_{k\to i})\big(\vec{\epsilon}_{k\to i}\cdot\hat{h}_{k\to i}\big)
\hat{h}_{k\to i} + 
b(h_{k\to i})\big[
\vec{\epsilon}_{k\to i} - \big(\vec{\epsilon}_{k\to i}\cdot\hat{h}_{k\to i}\big)\hat{h}_{k\to i}\big]\ ,
\label{eq:perturbations}
\end{equation}
where 
\begin{equation}
\begin{aligned}
a(h) = \frac{d\tilde{u}(h)}{dh} &= 
\frac{f(\beta J)f'(\beta h)}{f'\big\{f^{-1}[f(\beta J)f(\beta h)\big\}} = 
\frac{f(\beta J)f'(\beta h)}{f'(\beta\tilde{u})}\ ,\\
b(h) = \frac{\tilde{u}(h)}{h} &= 
\frac{f^{-1}[f(\beta J)f(\beta h)]}{\beta h}\ ,
\end{aligned}
\label{eq:AandB}
\end{equation}
and we have dropped the explicit dependence of $a(h)$ and $b(h)$ 
from $\beta$ and $J$ for simplicity. To elucidate the meaning of 
Eq.~\eqref{eq:perturbations} let us introduce the longitudinal 
$\mathcal{L}_{k\to i}$ 
and transverse projectors $\mathcal{T}_{k\to i}$ defined as 
\begin{equation}
\begin{aligned}
\mathcal{L}_{k\to i}^{\mu\nu} &= 
\frac{h^{\mu}_{k\to i} h^{\nu}_{k\to i}}{|h_{k\to i}|^2}\ ,\\
\mathcal{T}_{k\to i}^{\mu\nu} &= \delta_{\mu\nu} - \frac{h^{\mu}_{k\to i} h^{\nu}_{k\to i}}{|h_{k\to i}|^2}\ ,
\end{aligned}
\label{eq:projectors}
\end{equation}
where $\mathcal{L}_{k\to i}$ is a $n\times n$ matrix that 
projects an arbitrary vector on the direction parallel to 
$\vec{h}_{k\to i}$,  while $\mathcal{T}_{k\to i}$ projects 
on the $(n-1)$-dimensional subspace orthogonal to the cavity 
field. Using the projectors defined in Eq.~\eqref{eq:projectors} 
we can rewrite Eq.~\eqref{eq:perturbations} as 
\begin{equation}
\vec{\epsilon}_{i\to j} = \sum_{k\in\partial i\setminus j}
\Big[a(h_{k\to i})\mathcal{L}_{k\to i} + 
b(h_{k\to i})\mathcal{T}_{k\to i}\Big]
\vec{\epsilon}_{k\to i}\ .
\label{eq:perturbations2}
\end{equation}
At this point we can introduce the $2Mn\times 2Mn$
stability matrix $\mathcal{M}$ defined on the $2M$ 
directed edges of the graph as 
\begin{equation}
\mathcal{M}_{i\to j, k\to l}^{\mu\nu} = 
\Big[
a(h_{k\to l})\mathcal{L}_{k\to l}^{\mu\nu} + 
b(h_{k\to l})\mathcal{T}_{k\to l}^{\mu\nu}
\Big]\mathcal{B}_{i\to j, k\to l}\ ,
\label{eq:SIstabilityMatrix}
\end{equation}
where $\mathcal{B}$ is the {\it non-backtracking} matrix of the 
graph~\cite{hashimoto} of size $2M\times 2M$, that has nonzero 
entries only when $k\to l, i\to j$ form a pair of consecutive 
non-backtracking directed edges, i.e. when $l=i$ and $k\neq j$. 
By means of $\mathcal{M}$ we can rewrite Eq.~\eqref{eq:perturbations2} 
in the following compact form 
\begin{equation}
\vec{\mathcal{E}} = \mathcal{M}\vec{\mathcal{E}}\ ,    
\end{equation}
where $\vec{\mathcal{E}}$ is a vector with $2Mn$ entries obtained by 
column staking the $2M$ vectors $\vec{\epsilon}_{i\to j}$. 
Eigenvalues of the stability matrix $\mathcal{M}$ fully determine 
the fate of an arbitrary perturbation $\vec{\mathcal{E}}$ or, 
equivalently, the stability of the fixed point solution. 
Specifically, stability of the solution requires that the maximum 
eigenvalue $\lambda_1\leq 1$. Moreover, eigenvectors of $\mathcal{M}$ 
give a complete description of the collective fluctuations (normal 
modes) around the fixed point. 
To familiarize with the stability matrix, let us first consider 
the case of a pure ferromagnetic model without external field.
In this case, due to the homogeneity of the connectivity of the 
random regular graph, the cavity equations admits the homogeneous 
solution $\vec{h}_{i\to j}=\vec{h}$ for all directed edges $\{i\to j\}$. 
As a consequence, the factor in square brackets in Eq.~\eqref{eq:SIstabilityMatrix} 
is decoupled from the non-backtracking matrix and the stability matrix 
reduces to the tensor product form 
\begin{equation}
\mathcal{M} = \Big[a(h)\mathcal{L} + b(h)\mathcal{T}\Big]\otimes\mathcal{B}\ ,
\label{eq:SIstabilityMatrix2}
\end{equation}
whose eigenvalues $\lambda_{\mathcal{M}}$ are simply related to the 
eigenvalues of the non-backtracking matrix $\lambda_{\mathcal{B}}$ and 
to the coefficients $a(h)$ and $b(h)$ through 
\begin{equation}
\lambda_{\mathcal{M}} = 
\left\{
\begin{matrix}
a(h)\lambda_{\mathcal{B}} \equiv \lambda_L\\
b(h)\lambda_{\mathcal{B}} \equiv \lambda_T
\end{matrix}
\right. \ .
\end{equation}
Eigenvalues $\lambda_L$ and $\lambda_T$ describe the rate of decay of 
perturbations longitudinal and transverse to the cavity field, respectively. 
It is easy to see that matrix $\mathcal{M}$ in Eq.~\eqref{eq:SIstabilityMatrix2} 
has two types of eigenvectors: $2M$ longitudinal eigenvectors of the form 
$V_L=\vec{h}\otimes|B\rangle$ (where $|B\rangle$ is the eigenvector of the 
non-backtracking matrix) with eigenvalues $\lambda_L$ that describe the 
longitudinal fluctuations; and $2M(n-1)$ transverse eigenvectors of 
the form $V_T=\vec{v}^{\ a}_\perp\otimes|B\rangle$, for $a=1,...,n-1$ 
(where $\{\vec{v}^{\ a}_\perp\}$ span the subspace orthogonal to $\vec{h}$) 
with eigenvalues $\lambda_T$ describing the behavior of the transverse 
fluctuations. 
In the paramagnetic phase $\vec{h}=0$ and we find $a(0)=b(0)=f(\beta J)$, 
so that the two eigenvalues are degenerate, i.e. $\lambda_L = \lambda_T$. 
The phase transition occurs at the point where the solution 
$\vec{h}=0$ becomes unstable, i.e. when $f(\beta_c J)\lambda_{\mathcal{B}}=1$. 
Choosing the largest eigenvalue of the non-backtracking matrix, 
$\lambda_{\mathcal{B}} = C-1$, we obtain the following analytic 
expression for the critical temperature of the pure model:
\begin{equation}
\frac{T_c}{J} = 
\frac{1}{f^{-1}\left(\frac{1}{C-1}\right)}\ ,
\label{eq:Tc}
\end{equation}
which agrees with the known results for $n=1,2$~\cite{montanari,castillo}.  
Below $T_c$, in the ferromagnetic phase, the cavity field is 
non-zero, $\vec{h}\neq0$, and the degeneracy between the two 
eigenvalues $\lambda_L, \lambda_T$ is lifted since $a(h)\neq b(h)$. 
There are two types of fluctuations: a longitudinal one 
along the direction of the cavity field, and $n-1$ transverse 
ones in the $n-1$ directions perpendicular to the cavity field. 
To understand the stability of the ferromagnetic solution we 
have to understand how a perturbation applied to the fixed 
point solution evolves under subsequent iterations of the 
cavity equations. We have to distinguish between a longitudinal 
and a transverse perturbation. A longitudinal perturbation 
$\vec{\epsilon}_L(0)$ at time $t=0$ will evolve, after $t$ 
iterations, as $\vec{\epsilon}_L(t) = [(C-1)a(h)]^t \vec{\epsilon}_L(0)$. 
Since $(C-1)a(h)<1$, longitudinal perturbations will eventually 
decay to zero, meaning that the solution is stable along 
the direction of the cavity field. The longitudinal perturbation 
is analogous to the Higgs mode of a ``Mexican hat'' potential.
On the other hand, transverse fluctuations evolve as 
$\vec{\epsilon}_T(t) = [(C-1)b(h)]^t \vec{\epsilon}_T(0)$. 
Since $(C-1)b(h)=1$, transverse perturbations that change 
the orientation of the cavity field will not decay 
to zero, meaning that the solution is only marginally stable 
along any direction orthogonal to the cavity field. 
Transverse perturbations are the Goldstone modes, also 
called spin waves.

\begin{figure}[h!]
\includegraphics[width=0.95\textwidth]{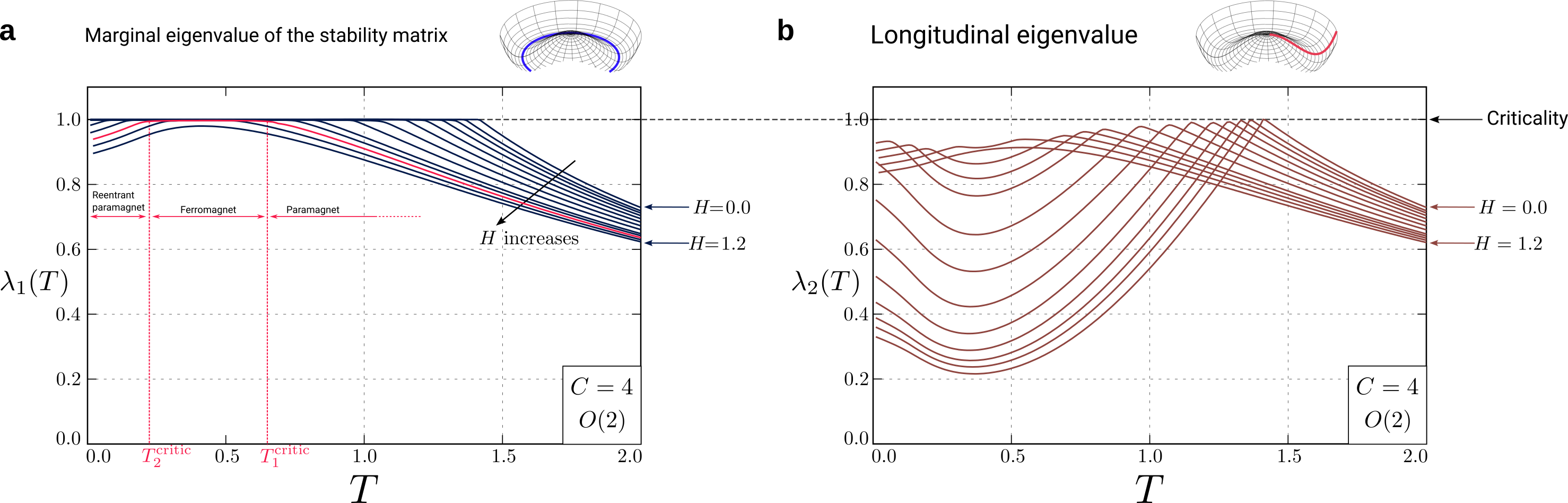} 
\caption{Marginal ({\bf a}) and longitudinal ({\bf b}) eigenvalue 
of the stability matrix~\eqref{eq:SIstabilityMatrix} of the 
RF $O(2)$ model on a random regular graph of connectivity 
$C=4$ and size $N=2\times 10^6$ for several values of the random 
field strength $H_R$. 
}
\label{fig:eigenvaluesSI}
\end{figure}

Having discussed the pure case, we move next to study the 
stability of the case with the random field. 
In this case the fixed point solution to the cavity 
Eqs.~\eqref{eq:cavityEqs} is not homogeneous and the stability 
matrix cannot be written in the tensor product~\eqref{eq:SIstabilityMatrix2}.
This means that global perturbations cannot be understood just 
by looking at local ones on individual directed edges, and thus we have 
to diagonalize the full matrix given in Eq.~\eqref{eq:SIstabilityMatrix}. 
Although we can still distinguish between longitudinal and transverse 
perturbations of the cavity fields locally on single edges, this 
distinction does not make sense at the macroscopic level, since 
global perturbations, described by the eigenvectors of the 
stability matrix, are a hybridization of local longitudinal 
and transverse modes. 
Having made this remark, we denote as $\vec{\mathcal{E}}$ 
the leading eigen-perturbation of the stability matrix and name 
it ``marginal perturbation'', since it generalizes the Goldstone 
mode of the simple ferromagnetic case. 
To compute the marginal perturbation, and its corresponding 
eigenvalue $\lambda_1$, we iterate Eq.~\eqref{eq:perturbations2} 
and normalize $\vec{\mathcal{E}}$ 
at each step as $\vec{\mathcal{E}}_{t+1} = \mathcal{M}\vec{\mathcal{E}}_t/|\mathcal{M}\vec{\mathcal{E}}_t|$. This way, $\vec{\mathcal{E}}_t$ 
converges to the dominant eigenvector and we obtain the 
largest eigenvalue by computing the Rayleigh quotient:
\begin{equation}
\lambda_1 = \lim_{t\to\infty}
\frac{\vec{\mathcal{E}}_t \cdot (\mathcal{M} \vec{\mathcal{E}}_t)}
{\vec{\mathcal{E}}_t \cdot \vec{\mathcal{E}}_t}\ , 
\label{eq:marginalMode}
\end{equation}
shown in Fig.~\ref{fig:eigenvaluesSI}a
To better understand the collective fluctuations we compute 
also the second leading eigenvector $\vec{\Delta}$ and 
its eigenvalue $\lambda_2$.  
The second eigen-perturbation is also a hybrid of longitudinal 
and transverse local fluctuations. That been said, we name it 
``longitudinal perturbation'', since it reduces to the canonical 
longitudinal mode in absence of random field. 
To compute $\vec{\Delta}$, the idea is to iteratively apply 
$\mathcal{M}$ to a vector belonging to the subspace orthogonal 
to $\vec{\mathcal{E}}$. 
In other words, we look for vectors $\vec{\delta}_{i\to j}$ 
such that $\vec{\delta}_{i\to j}\cdot \vec{\epsilon}_{i\to j}=0$ 
for all directed edges. Therefore, we introduce the projector 
$\mathcal{P}_{i\to j}$ defined as 
\begin{equation}
\mathcal{P}_{i\to j}^{\mu\nu} = \delta_{\mu\nu} - 
\frac{\epsilon^{\mu}_{i\to j} \epsilon^{\nu}_{i\to j}}
{|\epsilon_{i\to j}|^2}\ ,
\end{equation}
by means of which we can write down the iterative equations 
for $\vec{\delta}_{i\to j}$ as 
\begin{equation}
\vec{\delta}_{i\to j} = \sum_{k\in\partial i\setminus j}
\Big[a(h_{k\to i})\mathcal{P}_{i\to j}\mathcal{L}_{k\to i} + 
b(h_{k\to i})\mathcal{P}_{i\to j}\mathcal{T}_{k\to i}\Big]
\vec{\delta}_{k\to i}\ ,
\label{eq:perturbations3}
\end{equation}
which can be rewritten in a more explicit, although less compact, 
equivalent form as follows
\begin{equation}
\vec{\delta}_{i\to j} = \sum_{k\in\partial i\setminus j}
(a_k - b_k)\big(\vec{\delta}_{k\to i}\cdot\hat{h}_{k\to i}\big)
\Big[\hat{h}_{k\to i} - \hat{\epsilon}_{i\to j}
\big(\hat{\epsilon}_{i\to j}\cdot\hat{h}_{k\to i}\big)\Big] + 
b_k\Big[\vec{\delta}_{k\to i} - \hat{\epsilon}_{i\to j}
\big(\hat{\epsilon}_{i\to j}\cdot\vec{h}_{k\to i}\big)\Big]\ ,
\label{eq:perturbations4}
\end{equation}
where $a_k$ and $b_k$ are shorthand for $a(h_{k\to i})$ 
and $b(h_{k\to i})$, respectively. 
The longitudinal perturbation $\vec{\Delta}$ is obtained by 
stacking the $2M$ vectors $\vec{\delta}_{i\to j}$ and the 
second largest eigenvalue $\lambda_2$ is computed by the 
Rayleigh quotient
\begin{equation}
\lambda_2 = \lim_{t\to\infty}
\frac{\vec{\Delta}_t \cdot (\mathcal{M} \vec{\Delta}_t)}
{\vec{\Delta}_t \cdot \vec{\Delta}_t}\ , 
\label{eq:longitudinaMode}
\end{equation}
shown in Fig.~\ref{fig:eigenvaluesSI}b.

\subsection{Correlation functions}
In this section we study the correlation functions 
and show that the decay rate of the disorder-averaged 
connected correlation function is precisely the largest 
eigenvalue of the stability matrix $\mathcal{M}$. 

Let us consider two spins in the graph, denote them 
$\vec{s}_i$ and $\vec{s}_j$. Since the graph is locally 
tree-like, and connected, there will be at most one path 
connecting sites $i$ and $j$ whose length we denote as 
$\ell\equiv|i-j|$ (in a connected graph the distance between 
two nodes is defined as the number of edges in the shortest 
path connecting those two nodes.) 
It is convenient to rename the two spins as $\vec{s}_0$ 
and $\vec{s}_\ell$. 
The connected correlation is defined as 
\begin{equation}
\mathcal{C}_\ell^{\mu\nu} = 
\langle s_0^{\mu} s_\ell^{\nu}\rangle - 
\langle s_0^{\mu}\rangle \langle s_\ell^{\nu}\rangle\ ,\ \ \ \mu,\nu=1,...,n\ ,
\label{eq:correlation}
\end{equation}
and where the angular brackets $\langle \cdot \rangle$ 
indicates the average over the Boltzmann distribution. 
In practice, $\mathcal{C}_\ell$ can be computed by taking 
the derivative of $\langle s_0^{\mu}\rangle$ with respect 
to a perturbation applied on $s_\ell$. 
\begin{figure}[h]
\includegraphics[width=0.95\textwidth]{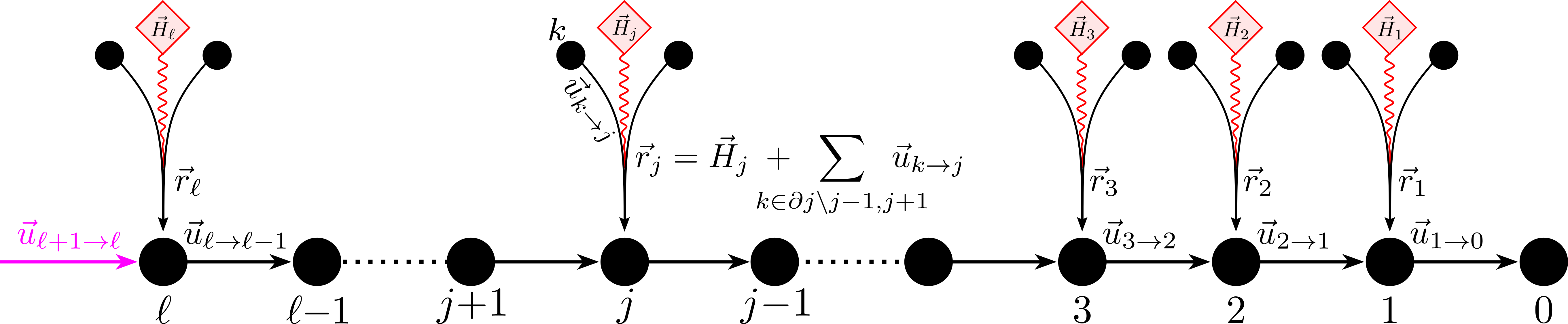} 
\caption{ Chain embedded in a random regular graph used 
to compute the correlation function between two spins at 
distance $\ell$ by means of Eq.~\eqref{eq:correlation}. 
The vector $\vec{r}_j$ is the sum of the random field on 
site $j$, namely $\vec{H}_j$, and the cavity biases $\vec{u}_{k\to j}$ 
coming from the $C-2$ branches of the graph outside the chain 
merging on site $j$ according to Eq.~\eqref{eq:cavityR}. 
}
\label{fig:chain}
\end{figure}
Using the visual representation in Fig.~\ref{fig:chain} 
it's easy to show that 
\begin{equation}
\mathcal{C}_\ell^{\mu\nu} = 
\frac{\partial\langle s_0^{\mu}\rangle}{\partial u^{\nu}_{\ell+1\to\ell}} = 
\sum_{\alpha_1,..., \alpha_\ell}
\frac{\partial\langle s_0^{\mu}\rangle}{\partial u^{\alpha_1}_{1\to0}}
\frac{\partial u^{\alpha_1}_{1\to0}}{\partial u^{\alpha_2}_{2\to1}}
\frac{\partial u^{\alpha_2}_{2\to1}}{\partial u^{\alpha_3}_{3\to2}}\dots 
\frac{\partial u^{\alpha_\ell}_{\ell\to\ell-1\to0}}{\partial u^{\nu}_{\ell+1\to\ell}}\ ,
\label{eq:correlation}
\end{equation}
The generic term in the sum on the right-hand-side of 
Eq.~\eqref{eq:correlation} is explicitly given by 
\begin{equation}
\begin{aligned}
\frac{\partial u^{\alpha}_{j\to j-1}}{\partial u^{\beta}_{j+1\to j}} &= 
\frac{\partial }{\partial u^{\beta}_{j+1\to j}}
\Bigg[\tilde{u}(\beta, J, |\vec{r}_j + \vec{u}_{j+1\to j}|)
\frac{r_j^\alpha + u^\alpha_{j+1\to j}}{|\vec{r}_j + \vec{u}_{j+1\to j}|}\Bigg] =\\
&=  
\frac{f(\beta J) f'(\beta|\vec{h}_{j\to j-1}|)}{f'(\beta|\vec{u}_{j\to j-1}|)}
\frac{h_{j\to j-1}^\alpha h_{j\to j-1}^\beta}{|\vec{h}_{j\to j-1}|^2} + 
\frac{|\vec{u}_{j\to j-1}|}{|\vec{h}_{j\to j-1}|}\Bigg(\delta^{\alpha\beta} - 
\frac{h_{j\to j-1}^\alpha h_{j\to j-1}^\beta}{|\vec{h}_{j\to j-1}|^2}\Bigg) = \\
&= 
a(h_{j\to j-1})\mathcal{L}_{j\to j-1}^{\alpha\beta} + 
b(h_{j\to j-1})\mathcal{T}_{j\to j-1}^{\alpha\beta}\ ,
\end{aligned}
\label{eq:dudu}
\end{equation}
where $\vec{r}_j$ is defined as 
\begin{equation}
\vec{r}_j \equiv \vec{h}_{j\to j-1} - \vec{u}_{j+1\to j} = 
\vec{H}_j\ +\hspace{-0.4cm}
\sum_{k\in\partial j\setminus j-1,j+1}
\hspace{-0.4cm}
\vec{u}_{k\to j}\ .
\label{eq:cavityR}
\end{equation}
Comparing Eq.~\eqref{eq:dudu} with Eq.~\eqref{eq:SIstabilityMatrix} we 
discover that 
\begin{equation}
\frac{\partial u^{\alpha}_{j\to j-1}}{\partial u^{\beta}_{j+1\to j}} = 
\mathcal{M}_{j+1\to j, j\to j-1}^{\alpha\beta}\ ,
\end{equation}
where $\mathcal{M}$ is the stability matrix. 
Since the large distance decay of the correlation function 
is determined by the behavior of the product of derivatives 
in Eq.~\eqref{eq:correlation} we may equally consider the 
following definition of the correlation function
\begin{equation}
\begin{aligned}
\mathcal{C}_\ell^{\alpha_1\alpha_{\ell+1}} &= 
\sum_{\alpha_2,..., \alpha_\ell}\prod_{k=1}^{\ell}
\frac{\partial u^{\alpha_k}_{k\to k-1}}{\partial u^{\alpha_{k+1}}_{k+1\to k}}\ ,\\ 
u^{\alpha_k}_{k\to k-1} &= 
\tilde{u}(\beta, J, |\vec{r}_k + \vec{u}_{k+1\to k}|)
\frac{\vec{r}^{\ \alpha_k}_k + \vec{u}^{\ \alpha_k}_{k+1\to k}}{|\vec{r}_k 
+ \vec{u}_{k+1\to k}|}\ .
\end{aligned}
\label{eq:lyapunov}
\end{equation}
This form suggests the following iterative equation for 
the correlation~\cite{moroneLDF} 
\begin{equation}
\begin{aligned}
\mathcal{C}_{\ell+1}^{\alpha\beta} &= 
\sum_{\gamma}\frac{\partial u^{\alpha}_{1\to0}}{\partial u^{\gamma}_{2\to1}}
\mathcal{C}_{\ell}^{\gamma\beta}\ ,\\
u^{\alpha}_{1\to0} &= 
\tilde{u}(\beta, J, |\vec{r} + \vec{u}_{2\to 1}|)
\frac{r^{\alpha} + u^{\alpha}_{2\to1}}{|\vec{r} + \vec{u}_{2\to 1}|}\ .
\end{aligned}
\end{equation}
We interpret these equations as a distributional equation 
for the joint probability $P_{\ell}(\mathcal{C},\vec{u})$, that 
reads
\begin{equation}
P_{\ell+1}(\mathcal{C},\vec{u}) = \mathbb{E}_r
\int d\mathcal{C}'d\vec{u}'\ P_{\ell}(\mathcal{C}',\vec{u}')
\prod_{\alpha\beta}\delta\Bigg[
\mathcal{C}_{\alpha\beta} - 
\sum_{\gamma}\frac{\partial u_{\alpha}}{\partial u'_{\gamma}}
\mathcal{C}'_{\gamma\beta}\Bigg]
\prod_{\alpha}\delta\Bigg[u_{\alpha} - 
\tilde{u}(\beta, J, |\vec{r} + \vec{u}'|)
\frac{r_{\alpha} + u'_{\alpha}}{|\vec{r} + \vec{u}'|}\Bigg]\ ,
\label{eq:PCu}
\end{equation}
where the expectation in front of the integral is taken with 
respect to the field $\vec{r}$ distributed with a $P(\vec{r})$ 
given by 
\begin{equation}
P(\vec{r}) = \mathbb{E}_{H}\int \Bigg[\prod_{k=1}^{C-2}d\vec{u}_k\ Q(\vec{u}_k)\Bigg]
\prod_{\alpha}\delta\Bigg[r^{\alpha} - H^{\alpha} - 
\sum_{k=1}^{C-2}u_k^{\alpha}\Bigg]\ . 
\end{equation}
Next, it is convenient to introduce the partial average 
\begin{equation}
\Psi^{\alpha\beta}_{\ell}(\vec{u}) = 
\int d\mathcal{C}\ P_{\ell}(\mathcal{C},\vec{u})\ \mathcal{C}^{\alpha\beta}\ ,
\end{equation}
whose meaning can be grasped by noticing that 
\begin{equation}
\int d\vec{u}\ \Psi^{\alpha\beta}_{\ell}(\vec{u}) = 
\int d\vec{u}\ d\mathcal{C}\ P_{\ell}(\mathcal{C},\vec{u})\ \mathcal{C}^{\alpha\beta} =\  
\overline{\mathcal{C}_{\ell}^{\alpha\beta}} =\  
\overline{\langle s_0^{\alpha} s_\ell^{\beta}\rangle} - 
\overline{\langle s_0^{\alpha}\rangle\langle s_\ell^{\beta}\rangle}\ ,
\end{equation}
where the overline, $\overline{\cdot}$, denotes average over the 
disorder (i.e. over the random graph and the random field). 
Multiplying Eq.~\eqref{eq:PCu} by $\mathcal{C}^{\mu\nu}$ on 
both sides and integrating in $d\mathcal{C}$ we obtain 
\begin{equation}
\Psi^{\mu\nu}_{\ell+1}(\vec{u}) = \mathbb{E}_r\int d\vec{u}'\ 
\delta\Bigg[\vec{u} - \tilde{u}(\beta, J, |\vec{r} + \vec{u}'|)
\frac{\vec{r} + \vec{u'}}{|\vec{r} + \vec{u}'|}\Bigg]
\sum_{\gamma}\frac{\partial u_{\mu}}{\partial u'_{\gamma}}
\Psi^{\gamma\nu}_{\ell}(\vec{u}')\ .
\label{eq:Psimunu}
\end{equation}
Now we assume that for $\ell\to\infty$ the functions 
$\Psi^{\alpha\beta}_{\ell}(\vec{u})$ decay exponentially 
and we set 
\begin{equation}
\begin{aligned}
\Psi^{\alpha\alpha}_{\ell}(\vec{u}) &\sim e^{-\gamma_1 \ell}G^{\alpha\alpha}(\vec{u})\ ,\\
\Psi^{\alpha\beta}_{\ell}(\vec{u}) &\sim e^{-\gamma_2 \ell}R^{\alpha\beta}(\vec{u})\ ,\ \ {\rm for}\ \alpha\neq\beta\ ,
\end{aligned}
\end{equation}
with $\gamma_1\geq 0$, $\gamma_2>0$, and the functions $G$ and $R$ 
normalized as 
\begin{equation}
\begin{aligned}
\int d\vec{u}\ G^{\alpha\alpha}(\vec{u}) &= 1\ ,\\
\int d\vec{u}\ R^{\alpha\beta}(\vec{u}) &= 1\ .
\end{aligned}
\end{equation}
The equation for $G^{\alpha\alpha}(\vec{u})$ reads 
\begin{equation}
e^{-\gamma_1}G^{\alpha\alpha}(\vec{u}) = \mathbb{E}_r\int d\vec{u}'\ 
\delta\Bigg[\vec{u} - \tilde{u}(\beta, J, |\vec{r} + \vec{u}'|)
\frac{\vec{r} + \vec{u'}}{|\vec{r} + \vec{u}'|}\Bigg]
\Bigg[\frac{\partial u_{\alpha}}{\partial u'_{\alpha}}
G^{\alpha\alpha}(\vec{u}') + 
e^{-(\gamma_2 - \gamma_1)\ell}\sum_{\beta\neq\alpha}\frac{\partial u_{\alpha}}{\partial u'_{\beta}}R^{\beta\alpha}(\vec{u}')\Bigg]\ .
\end{equation}
Next we suppose that the off-diagonal correlations decay 
faster than the diagonal ones, so that $\gamma_2>\gamma_1$, 
and we obtain an equation involving $G^{\alpha\alpha}$ only, 
which has the form of a Fredholm’s integral equation
\begin{equation}
\boxed{\ 
e^{-\gamma_1}G^{\alpha\alpha}(\vec{u}) = \mathbb{E}_r\int d\vec{u}'\ 
\delta\Bigg[\vec{u} - \tilde{u}(\beta, J, |\vec{r} + \vec{u}'|)
\frac{\vec{r} + \vec{u'}}{|\vec{r} + \vec{u}'|}\Bigg]
\frac{\partial u_{\alpha}}{\partial u'_{\alpha}}G^{\alpha\alpha}(\vec{u}')\ }\ ,
\label{eq:fredolm}
\end{equation}
On the other hand, taking the average over the disorder 
in Eq.~\eqref{eq:lyapunov} we find that 
\begin{equation}
\overline{\mathcal{C}_\ell^{\alpha\alpha}} = 
\sum_{\alpha_2,..., \alpha_\ell}
\overline{\prod_{k=1}^{\ell}
\mathcal{M}_{k+1\to k, k\to k-1}^{\alpha_k\alpha_{k+1}}}
\sim e^{-\ell \gamma_1}\ ,
\label{eq:prodRandomMat}
\end{equation}
where $\alpha_1=\alpha_{\ell+1}\equiv\alpha$. Most importantly, 
$\gamma_1$ in Eq.~\eqref{eq:prodRandomMat} is the Lyapunov 
exponent of the product of correlated random matrices 
$\mathcal{M}_{k+1\to k, k\to k-1}$, from which we infer that 
\begin{equation}
e^{-\gamma_1} = \frac{\lambda_1}{C-1}\ ,
\end{equation}
where $\lambda_1$ is precisely the largest eigenvalue of the 
stability matrix $\mathcal{M}$ defined in Eq.~\eqref{eq:marginalMode}. 
Integrating over $d\vec{u}$ in Eq.~\eqref{eq:fredolm} we find 
an analytic expression for $\lambda_1$ as
\begin{equation}
\boxed{\ 
\lambda_1 = (C-1)\mathbb{E}_r\int d\vec{u}\ 
G^{\alpha\alpha}(\vec{u})
\frac{\partial}{\partial u_{\alpha}}\Bigg[
\tilde{u}(\beta, J, |\vec{r} + \vec{u}|)
\frac{r_{\alpha} + u_{\alpha}}{|\vec{r} + \vec{u}|}\Bigg]\ }\ .
\end{equation}
Knowledge of $\lambda_1$ allows us to compute the ferromagnetic 
susceptibility
\begin{equation}
\chi_F = \frac{1}{N}\sum_{ij}
\Big[
\overline{\langle \vec{s}_i\cdot \vec{s}_j\rangle}-
\overline{\langle \vec{s}_i\rangle\cdot \langle \vec{s}_j\rangle}\Big]\propto  
\sum_{\ell}(C-1)^\ell
\Big[
\overline{\langle \vec{s}_0\cdot \vec{s}_\ell\rangle}-
\overline{\langle \vec{s}_0\rangle\cdot \langle \vec{s}_\ell\rangle}\Big]
\sim \frac{1}{1-\lambda_1}\ ,
\end{equation}
which diverges when $\lambda_1=1$. Moreover, a value of $\lambda_1>1$ 
is not physically acceptable and should be construed as a breakdown 
of the replica symmetric cavity method, as discussed next in Sec.~\ref{si:spinglass}. 
Before moving on, we conclude this section by deriving the 
equation for the off-diagonal correlation functions $R^{\alpha\beta}(\vec{u})$ 
and the decay rate $\gamma_2$. To this end, let us consider 
Eq.~\eqref{eq:Psimunu} for $\mu\neq\nu$, thus obtaining 
\begin{equation}
\begin{aligned}
e^{-(\gamma_2 - \gamma_1)\ell}e^{-\gamma_2}R^{\alpha\beta}(\vec{u}) &= 
\mathbb{E}_r\int d\vec{u}'\ 
\delta\Bigg[\vec{u} - \tilde{u}(\beta, J, |\vec{r} + \vec{u}'|)
\frac{\vec{r} + \vec{u'}}{|\vec{r} + \vec{u}'|}\Bigg]
\frac{\partial u_{\alpha}}{\partial u'_{\beta}}
G^{\beta\beta}(\vec{u}')\ \ +\\
&+
e^{-(\gamma_2 - \gamma_1)\ell}
\mathbb{E}_r\int d\vec{u}'\ 
\delta\Bigg[\vec{u} - \tilde{u}(\beta, J, |\vec{r} + \vec{u}'|)
\frac{\vec{r} + \vec{u'}}{|\vec{r} + \vec{u}'|}\Bigg]
\sum_{\gamma\neq\beta}\frac{\partial u_{\alpha}}{\partial u'_{\gamma}}R^{\gamma\beta}(\vec{u}')\ .
\end{aligned}
\label{eq:off-diagR}
\end{equation}
By letting $\ell$ tend to infinity in the previous equation we 
discover that 
\begin{equation}
\mathbb{E}_r\int d\vec{u}'\ 
\delta\Bigg[\vec{u} - \tilde{u}(\beta, J, |\vec{r} + \vec{u}'|)
\frac{\vec{r} + \vec{u'}}{|\vec{r} + \vec{u}'|}\Bigg]
\frac{\partial u_{\alpha}}{\partial u'_{\beta}}
G^{\beta\beta}(\vec{u}') = 0\ ,\ \ {\rm for}\ \ \alpha\neq \beta\ ,
\end{equation}
which, reinserted back into Eq.~\eqref{eq:off-diagR}, leads 
us to the self-consistent equation satisfied by $R^{\alpha\beta}(\vec{u})$:
\begin{equation}
\boxed{\ 
e^{-\gamma_2}R^{\alpha\beta}(\vec{u}) = \mathbb{E}_r\int d\vec{u}'\ 
\delta\Bigg[\vec{u} - \tilde{u}(\beta, J, |\vec{r} + \vec{u}'|)
\frac{\vec{r} + \vec{u'}}{|\vec{r} + \vec{u}'|}\Bigg]
\sum_{\gamma\neq\beta}\frac{\partial u_{\alpha}}{\partial u'_{\gamma}}R^{\gamma\beta}(\vec{u}')\ }\ .
\label{eq:off-diagR2}
\end{equation}
Finally, setting 
\begin{equation}
e^{-\gamma_2} = \frac{\lambda_2}{C-1}\ ,
\end{equation}
and integrating over $d\vec{u}$ in Eq.~\eqref{eq:off-diagR2} 
we find an analytic expression for $\lambda_2$, given by 
\begin{equation}
\boxed{\ 
\lambda_2 = (C-1)\mathbb{E}_r\int d\vec{u}\ 
\sum_{\gamma\neq\beta}R^{\gamma\beta}(\vec{u})
\frac{\partial}{\partial u_{\gamma}}\Bigg[
\tilde{u}(\beta, J, |\vec{r} + \vec{u}|)
\frac{r_{\alpha} + u_{\alpha}}{|\vec{r} + \vec{u}|}\Bigg]\ }\ .
\end{equation}

\section{Spin-glass model}
\label{si:spinglass}
For completeness we solved the cavity equations for 
a spin glass model on a random regular graph with 
Hamiltonian given by
\begin{equation}
\mathcal{H} = -\frac{1}{2}\sum_{ij}J_{ij}A_{ij}\vec{s}_i\cdot\vec{s}_j\ ,
\label{eq:spinglass}
\end{equation}
where $J_{ij}$ are normally distributed with zero mean 
and unit variance. 
The main reason to consider this model in this work is 
to run a sanity check on the ability of the eigenvalue 
$\lambda_1(T)$ to effectively detect a spin-glass phase with 
replica symmetry breaking. In fact, the model described 
by the Hamiltonian in Eq.~\eqref{eq:spinglass} is believed  
to have a spin-glass phase and thus is a good model to 
establish whether the RSB instability can be determined by 
computing the largest eigenvalue $\lambda_1$ of the stability 
matrix, as discussed next. 

\begin{figure}[h!]
\includegraphics[width=0.71\textwidth]{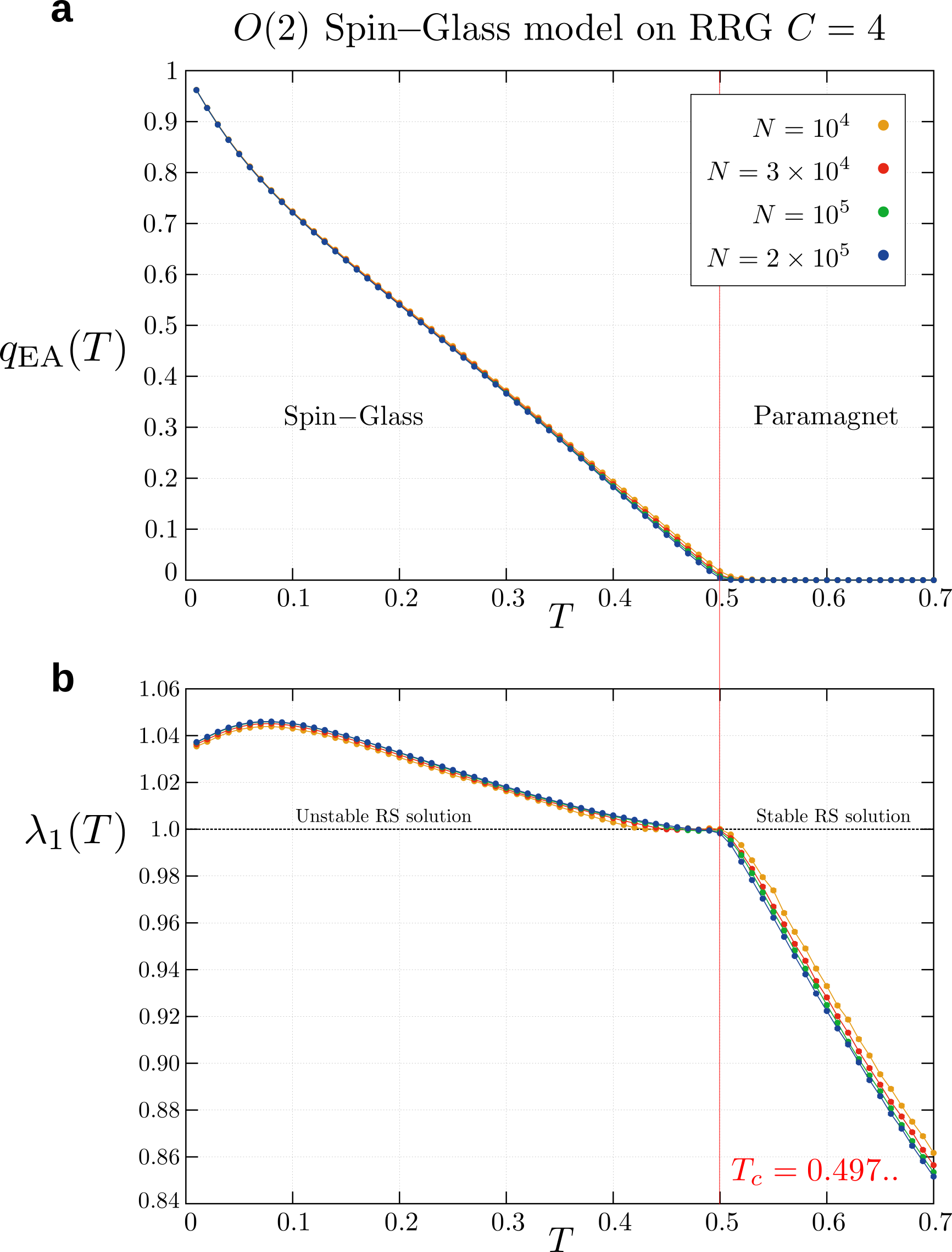} 
\caption{{\bf Spin glass $O(2)$ model on a RRG of connectivity $C=4$.}
{\bf a}, Edwards-Anderson order parameter $q_{EA}(T)$ for a spin-glass 
model with Gaussian random couplings on a RRG of degree $C=4$ and $n=2$. 
Different curves correspond to different system sizes ranging from 
$N=10^4$ to $N=2\times10^5$. Each curve is averaged over 100 samples 
(error bars are smaller than symbol size). 
We find a phase transition from a paramagnetic phase, where $q_{EA}=0$, 
to a spin-glass phase, where $q_{EA}\neq0$, at a temperature $T_c = 0.497...$ 
given by the solution to Eq.~\eqref{eq:Tc-spinglass}. 
{\bf b}, Largest eigenvalue of the stability matrix showing that the 
replica-symmetric solution is stable above $T_c$ but unstable below. 
The curves correspond to the same system's sizes as in {\bf a} averaged 
over 100 realizations of the couplings and the random graphs (error bars 
are smaller than symbol size). 
}
\label{fig:sgrrg}
\end{figure}

First of all we obtained the critical temperature analytically 
for any $n$ and $C$ as the solution of the following equation:
\begin{equation}
\int_{-\infty}^{+\infty}\frac{dJ}{2\pi}e^{-J^2/2}f(\beta_cJ)^2 = \frac{1}{C-1}\ ,
\label{eq:Tc-spinglass}
\end{equation}
which for $n=2$ and $C=4$ gives 
\begin{equation}
T_c = 0.4972898...\ ,
\end{equation}
which marks the transition from a paramagnetic phase where 
$q_{EA}=0$, to a spin-glass phase, where $q_{EA}\neq0$, 
as seen in Fig.~\ref{fig:sgrrg}a. 
Figure~\ref{fig:sgrrg}b shows the largest eigenvalue 
$\lambda_1$ of the stability matrix as a function of $T$ for 
the same model with $n=2$ and $C=4$. This eigenvalue reaches 
$1$ at the critical temperature and is larger than $1$ for 
$T<T_c$, meaning that the replica symmetric solution is always 
unstable in the spin-glass phase. 

A complete analysis of the spin-glass model defined by the 
Hamiltonian in Eq.~\eqref{eq:spinglass} along with the computation 
of the full de-Almeida-Thouless line will be published 
elsewhere~\cite{moroneselsTBP}. 

\section{Ground state, localization of low energy excitations, 
and screening of disorder}
\label{sec:groundstate}
In this section we study the ground state and the spectrum 
of the low energy excitations through a numerical minimization 
of the energy function and diagonalization of the corresponding 
Hessian. 
To be definite we focus to the case $n=2$. 
In this case the spin variables $\vec{s}_i$ can be 
represented by a single real number $\theta_i\in[0,2\pi)$. 
Calling $\bar{\theta}$ the ground state configuration, 
we can expand the energy function up to second order 
around $\bar{\theta}$ as 
\begin{equation}
E(\theta) = E(\bar{\theta}) + 
\frac{1}{2}\sum_{ij}(\theta_i - \bar{\theta}_i)
\mathcal{H}_{ij}(\bar{\theta})(\theta_j - \bar{\theta}_j)\ ,
\end{equation}
where $\mathcal{H}$ is the Hessian given by 
\begin{equation}
\mathcal{H}_{ij}(\theta) = \delta_{ij}\left[
|H_i|\cos(\theta_i-\phi_i) + 
\sum_{k=1}^NA_{ik}\cos(\theta_i-\theta_k)\right] - 
A_{ij}\cos(\theta_i-\theta_j)\ .
\end{equation}
We observe immediately that, in absence of external 
field, $H_i=0$, and since $\bar{\theta}_i = \bar{\theta}_j$ 
for all $i,j$, then the Hessian is simply given by 
the graph Laplacian, i.e. $\mathcal{H} = D-A$, where 
$D_{ij}= C\delta_{ij}$. 
The smallest eigenvalue of the graph Laplacian is identically 
zero, $e_0=0$, with multiplicity equal to the number of 
connected components of the graph, hence in our case the 
multiplicity is one since the random regular graph has only 
one connected component by construction.  
The corresponding eigenvector is the uniform vector 
given by $|v_0\rangle = \frac{1}{\sqrt{N}}(1,1,...,1)$, 
which represents the Goldstone mode. 
The second smallest eigenvalue is strictly positive,  
$e_1=C-2\sqrt{C-1}>0$, so the spectrum of the Hessian 
is gapped above the zero mode. 

Next, we consider the case of nonzero random field. 
First, we need an important ingredient: the participation 
ratio which quantifies the degree of localization of an 
eigenmode, defined as 
\begin{equation}
PR(\vec{v}) \equiv \frac{\langle v^2\rangle^2}{\langle v^4\rangle}=
\frac{\left(\sum_{i=1}^{N}v_i^2\right)^2}{N\sum_{i=1}^{N}v_i^4}\ .
\label{eq:partRatio}
\end{equation} 
Roughly speaking, when a vector is localized, only a $O(1)$ 
number of components are nonzero, and thus $PR(\vec{v})=O(1/N)\to 0$. 
On the contrary, a vector which is completely delocalized 
has $PR(\vec{v})=1$. For example the zero mode of the 
pure ferromagnetic model is delocalized over the whole 
graph, i.e. $PR(\vec{v}_0)=1$. 

When the random field is switched on we observe that 
the zero mode starts to localize, as signaled by the 
fact that $PR(\vec{v}_0)<1$ and shown in Fig.~\ref{fig:figHessiandiagram}a 
of the main text. Simultaneously, the gap shrinks as 
the random field increases, as seen in Fig.~\ref{fig:figHessiandiagram}b 
of the main text. 
At $H_R=H_c\sim 1$ the zero mode is fully localized and the 
spectrum becomes gapless.
Furthermore, we observe the appearance of a mobility edge, 
i.e., an interval of eigenvalues $[0, e_*]$ such that the 
participation ratio of all eigenvectors corresponding to 
eigenvalues in this interval (denoted $PR(e)$ with a slight 
abuse of notation) vanishes:
\begin{equation}
PR(e) = 0 \ \ \ {\rm for}\ e\in [e_0, e_*]\ .
\end{equation}

Now, let $H_R=H_c+\epsilon$ and let's analyze the effect 
of thermal fluctuations. To understand their effect we 
need to look at the $PR(e)$ as a function of the Hessian 
eigenvalues $e$. 
If the temperature is smaller than the mobility edge, $T<e_*$, 
only localized modes are excited and thus stays 
paramagnetic. 
However, when $T$ exceeds the mobility edge, $T>e_*$, 
thermal fluctuations can excite delocalized modes and 
the system magnetizes for all $H_R\in[H_c,H_{max}]$ 
at large enough $T$. This is the physical mechanism 
which explains the re-entrant phase transition occurring 
at finite temperature. 
Next we discuss the physical interpretation of the 
re-entrance in terms of the effective screening of 
the quenched disorder mediated by thermal fluctuations.

\subsection{Thermal screening of the random field}
\label{si:screening}
To be definite we consider the $n=2$ random field model 
described by the Hamiltonian 
\begin{equation}
H(\theta) = -\frac{1}{2}\sum_{i,j}J_{ij}\cos(\theta_i-\theta_j) 
-\sum_iH_i\cos(\theta_i-\phi_i)\ ,
\end{equation}
where, for the time being, we only require the couplings 
$J_{ij}$ to be symmetric, $J_{ij}=J_{ji}$, and $H_i$, 
the modulus of the local field, to be non-negative, 
$H_i\geq0$. 
Together with the model described by $H(\theta)$ we also 
consider an auxiliary (Gaussian) model described by the 
Hamiltonian $H_0(\theta)$ given by 
\begin{equation}
H_0(\theta) = \frac{1}{2}\sum_{i,j}
(\theta_i-\overline{\theta}_i)(\mathcal{C}^{-1})_{ij}(\theta_j-\overline{\theta}_j)\ ,
\end{equation}
where matrix $\mathcal{C}$ and vector $\overline{\theta}$ 
are variational parameters to be determined self-consistently, 
as explained next. Notice that $\mathcal{C}$ must be positive 
semidefinite in order for the auxiliary model to have a ground 
state energy bounded from below. 
The partition function of the original model can be written as 
\begin{equation}
Z = \int \prod_id\theta_i\ e^{-\beta H(\theta)} = 
Z_0\int \prod_id\theta_i\ \frac{e^{-\beta H_0(\theta)}}{Z_0}e^{-\beta (H-H_0)} = 
Z_0\Big\langle e^{-\beta (H-H_0)}\Big\rangle_0\ .
\end{equation}
Using the convexity of the exponential we have  
\begin{equation}
Z\geq Z_0e^{-\beta \langle H-H_0\rangle_0}\ ,
\end{equation}
or equivalently 
\begin{equation}
F\leq F_0 + \langle H-H_0\rangle_0\ ,
\end{equation}
which is nothing but the Jensen-Bogoliubov inequality. 
Since $\langle H_0\rangle_0$ does not depend on $\mathcal{C}$ 
and $\overline{\theta}$ it can be neglected. 
Denoting $E^{ren}(\mathcal{C}, \overline{\theta})\equiv\langle H\rangle_0$,   
we define the effective free energy as 
\begin{equation}
\Phi(\mathcal{C}, \overline{\theta}) = 
E^{ren}(\mathcal{C}, \overline{\theta}) - TS_0(\mathcal{C}) 
\ , 
\end{equation}
where $S_0(\mathcal{C}) = \frac{1}{2}\log\det\mathcal{C}$. 
The average $\langle H\rangle_0$ can be performed exactly 
and we obtain the following analytical expression of $\Phi$ 
(neglecting terms independent from $\mathcal{C}$ and $\overline{\theta}$)
\begin{equation}
\Phi(\mathcal{C}, \overline{\theta}) = -\frac{T}{2}\log\det\mathcal{C} 
-\frac{1}{2}\sum_{i,j}J_{ij}^{ren}\cos(\overline{\theta}_i-\overline{\theta}_j) 
-\sum_i H_i^{ren}\cos(\overline{\theta}_i-\phi_i)\ ,
\label{eqsi:varfreeEnergy}
\end{equation}
where the renormalized couplings and random fields are 
given by 
\begin{equation}
\begin{aligned}
J_{ij}^{ren} &= J_{ij}\exp\Big[-\frac{T}{2}
\Big(\mathcal{C}{ii}+\mathcal{C}_{jj}-2\mathcal{C}_{ij}\Big)\Big]\ ,\\
H_i^{ren} &= H_i \exp\Big(-\frac{T}{2}\mathcal{C}_{ii}\Big)\ .
\end{aligned}
\end{equation}
Note that not all $\mathcal{C}{ii}$ can be negative because the 
matrix must be positive semidefinite. Therefore whenever $\mathcal{C}{ii}>0$ 
the local random field gets screened at nonzero temperature and 
reduced by a factor $e^{-T\mathcal{C}_{ii}/2}$. 
Notice also the peculiar form of the screening factor, which is 
not in the canonical Arrhenius form. 
The partial derivative of $\Phi$ with respect to $\overline{\theta}_k$ 
is 
\begin{equation}
\frac{\partial \Phi}{\partial \overline{\theta}_k} = 
\frac{\partial E^{ren}}{\partial \overline{\theta}_k} = 
\sum_j J_{kj}^{ren}\sin(\overline{\theta}_k-\overline{\theta}_j) + 
H_k^{ren}\sin(\overline{\theta}_k-\phi_k)\ ,
\end{equation}
and the derivative with respect to $\mathcal{C}_{ij}$ is 
\begin{equation}
\frac{\partial \Phi}{\partial \mathcal{C}_{ij}} = 
-\frac{T}{2}(\mathcal{C}^{-1})_{ij} + 
\delta_{ij}\frac{T}{2}\Big[\sum_k J_{ik}^{ren}\cos(\overline{\theta}_i-\overline{\theta}_k) + 
H_i^{ren}\cos(\overline{\theta}_i-\phi_i)\Big] - 
\frac{T}{2}J_{ij}^{ren}\cos(\overline{\theta}_i-\overline{\theta}_j)\ .
\end{equation}
Setting to zero the partial derivatives of $\Phi$ we obtain the 
equations determining $\mathcal{C}$ and $\overline{\theta}$ given by 
\begin{equation}
\begin{aligned}
0 &= \frac{\partial E^{ren}}{\partial \overline{\theta}_i}\ ,\\
\mathcal{C}^{-1}_{ij} &= 
\frac{\partial^2 E^{ren}}{\partial \overline{\theta}_i \partial \overline{\theta}_j}\ ,
\end{aligned}
\end{equation}
where the last equation can be easily proved using the definition of 
$E^{ren}$, which is 
\begin{equation}
E^{ren}(\mathcal{C}, \overline{\theta}) = 
-\frac{1}{2}\sum_{i,j}J_{ij}^{ren}\cos(\overline{\theta}_i-\overline{\theta}_j) 
-\sum_i H_i^{ren}\cos(\overline{\theta}_i-\phi_i)\ .
\end{equation}

\end{document}